\documentclass[12pt]{iopart}

\usepackage{iopams}  
\usepackage{graphicx}
\usepackage{xcolor}
\usepackage{ulem}
\newcommand{\bq}{\begin{equation}}
\newcommand{\eq}{\end{equation}}
\newcommand{\bqn}{\begin{eqnarray}}
\newcommand{\eqn}{\end{eqnarray}}
\newcommand{\nb}{\nonumber}
\newcommand{\lb}{\label}

\graphicspath{{Figuras/}}
\definecolor{OliveGreen}{rgb}{0,0.6,0}

\begin{document}
	
\title{Cylindrical Gravitational Waves in Einstein-Aether Theory}

\author{R. Chan$^{1}$, M. F. A. da Silva$^{2}$ and V. H. Satheeshkumar$^{3}$} 

\address{
			$^{1}$Coordena\c{c}\~{a}o de Astronomia e Astrof\'{i}sica,
			Observat\'{o}rio Nacional (ON), Rio de Janeiro, RJ 20921-400, Brazil
			\\
			$^{2}$Departamento de F\'{i}sica Te\'{o}rica, 
			Universidade do Estado do Rio de Janeiro (UERJ), Rio de Janeiro, RJ 20550-900, Brazil
			\\
			$^{3}$Departamento de F\'{\i}sica, Universidade Federal do Estado do Rio de Janeiro (UNIRIO), Rio de Janeiro, RJ 22290-240, Brazil
			}
			
\ead{chan@on.br, mfasnic@gmail.com, vhsatheeshkumar@gmail.com}

\date{\today}

\begin{abstract}
Along the lines of the Einstein-Rosen wave equation of General Relativity (GR), we derive a gravitational wave equation with cylindrical symmetry in the Einstein-aether (EA) theory. We show that the gravitational wave in the EA is periodic in time for both the metric functions $\Psi(r,t)$ and $H(r,t)$. However, in GR, $\Psi(r,t)$ is periodic in time, but $H(r,t)$ is semi-periodic in time, having a secular drifting in the wave frequency. The evolution of wave pulses of a given width is entirely different in both theories in the $H(r,t)$ metric function due to this frequency drifting. Another fundamental difference between the two theories is the gravitational wave velocity. While in GR, the waves propagate with the speed of light,  in EA, there is no upper limit to the wave velocity, reaching infinity if $c_{13} \rightarrow 1$ and zero if $c_{13} \rightarrow -\infty$. We also show that energy-momentum pseudotensor and superpotential get contributions from aether in addition to the usual gravitational field part. All these characteristics are observational signatures that differentiate GR and EA.
\end{abstract}

\maketitle

\date{}

\def\beq{\begin{equation}}
	\def\eeq{\end{equation}}
\def\bea{\begin{eqnarray}}
	\def\eea{\end{eqnarray}}
\def\ben{\begin{enumerate}}
	\def\een{\end{enumerate}}
\def\la{\langle}
\def\ra{\rangle}
\def\a{\alpha}
\def\b{\beta}
\def\g{\gamma}\def\G{\Gamma}
\def\d{\delta}
\def\e{\epsilon}
\def\phi{\varphi}
\def\k{\kappa}
\def\l{\lambda}
\def\m{\mu}
\def\n{\nu}
\def\o{\omega}
\def\p{\pi}
\def\r{\rho}
\def\s{\sigma}
\def\t{\tau}
\def\L{{\cal L}}
\def\S{\Sigma }
\def\gsim{\; \raisebox{-.8ex}{$\stackrel{\textstyle >}{\sim}$}\;}
\def\lsim{\; \raisebox{-.8ex}{$\stackrel{\textstyle <}{\sim}$}\;}
\def\gtrsim{\gsim}
\def\lessim{\lsim}
\def\loc{{\rm local}}
\def\vm{v_{\rm max}}
\def\bh{\bar{h}}
\def\del{\partial}
\def\nab{\nabla}
\def\half{{\textstyle{\frac{1}{2}}}}
\def\fourth{{\textstyle{\frac{1}{4}}}}

\section{Introduction}

Immediately after completing the theory of General Relativity \cite{Einstein:1915ca} and proposing the three classical tests \cite{Einstein:1916vd} to verify the theory experimentally, in 1916, Einstein predicted the existence of gravitational waves as a natural consequence of dynamical spacetime \cite{Einstein:1916cc}. In this first article \cite{Einstein:1916cc} on gravitational waves, Einstein concluded that accelerated massive objects generate changes in the spacetime curvature, which propagate like waves with the speed of light. However, he concluded that gravitational waves carry no energy because of a calculational error, which he corrected in 1918 \cite{Einstein:1918btx}. He also derived a famous formula for the energy loss of a system emitting gravitational radiation. Nevertheless, Einstein tried to disprove the existence of gravitational waves in 1937 \cite{Einstein-Rosen}. The issue of gravitational waves remained controversial until Bondi \cite{Bondi:1957dt} settled it for good, which was later explored in a series of papers by Bondi and his collaborators. The first experimental evidence, albeit indirect, came in 1979 with the discovery of the orbital decay of the newly discovered binary pulsar by Hulse and Taylor \cite{Taylor:1979zz}. Finally, the gravitational waves were directly detected in 2015 by the LIGO and Virgo collaborations \cite{LIGOScientific:2016aoc}. Now, with this direct detection of gravitational waves, we have a new window into the universe and can test our understanding of gravity at higher energies like never before. A fascinating account of the history of gravitational waves can be found in references \cite{Kennefick, Buchwald, Will-Yunes}.

Einstein and Rosen were the first to consider cylindrical gravitational waves. They solved the field equations of the time-dependent vacuum with cylindrical symmetry \cite{Einstein-Rosen}. This solution described the outer spacetime for a collapsing cylindrical source, producing gravitational waves. In fact, cylindrical symmetry is the simplest symmetry capable of producing gravitational waves, as the Birkhoff theorem forbids the production of gravitational waves from spherical symmetry \cite{Chan:2022mxd}. A monochromatic wave, or a pulse of cylindrical symmetry, moves inward in matter-free space, implodes on the axis, and moves out again. This is the only problem of gravitational radiation where one has an exact solution to the general relativity field equations. Although this special solution does not illustrate all the salient features of gravitational radiation, we can still make correct general statements about gravitational waves. This problem, therefore, occupies a unique position in the theory of gravitational radiation. Also, the cylindrical spacetimes are of great interest, as they allow the study of a wide range of physical phenomena, from the gravitational analog of the Aharonov-Bohm effect to the cosmic superstrings. A recent review of this topic can be found in \cite{Bronnikov:2019clf}. The Einstein-Rosen waves are still investigated from very different perspectives, from the Renormalization group approach \cite{Harada:2013wap} to the Geroch group \cite{Penna:2021kua}; from chaos theory \cite{Szybka:2023kmx} to holography \cite{Penna:2023pvg}; and from gravitational collapse \cite{Garcia-Parrado:2018swu} to observable quantum gravity effects \cite{He:2022tvk}. More on cylindrical gravitational waves can be found in the recent articles \cite{Bini:2018shh} and \cite{Chen:2020ktv}.

Since the first observation of gravitational waves, several works have studied the gravitational waves in EA theory, for example, the references \cite{Oost:2018tcv}, \cite{Oost:2018oww}, \cite{Lin:2018ken} and \cite{Zhang:2019iim}. Even before that, Jacobson and Mattingly \cite{Jacobson2004} obtained the gravitational wave modes by linearizing the EA field equations.  
In a previous work, Chan \& da Silva (2021)  \cite{Chan2021} present all the possible solutions for a static cylindrical symmetric spacetime in the EA theory. In this work, we obtain the analog of the  Einstein-Rosen wave equations using the full EA field equations without the linearization.

The paper is organized as follows. In Section 2, we present the EA field equations. In Section 3, we compute field equations for a general cylindrical metric, obtain the cylindrical gravitational wave equation in Section 4, and solve it in Section 5. In Sections 6 and 7, we study the gravitational wave pulse. We compute energy-momentum pseudotensor and superpotential in Section 8. We summarize our results in Section 9.

\section{Field equations in EA theory }

Here, we give a very brief summary of the EA theory. We refer the reader to our earlier papers on the subject for more details \cite{Campista:2018gfi, Chan:2019mdn, Chan:2020amr, Chan:2021ela, Satheeshkumar2021}. The general action of the EA theory, in a background where the  metric signature is $({-}{+}{+}{+})$ and with the units chosen such that the speed of light defined by the metric $g_{ab}$ is unity, is given by 
\bq 
S = \int \sqrt{-g}~(L_{\rm Einstein}+L_{\rm aether}+L_{\rm matter}) d^{4}x,
\label{action}
\eq
where the first term is the usual Einstein-Hilbert Lagrangian, defined by $R$, the Ricci scalar, and $G$, the EA gravitational constant, as
\bq 
L_{\rm Einstein} =  \frac{1}{16\pi G} R. 
\eq
The second term, the aether Lagrangian, is given by
\bq 
L_{\rm aether} =  \frac{1}{16\pi G} [-K^{ab}{}_{mn} \nabla_a u^m
\nabla_b u^n +
\lambda(g_{ab}u^a u^b + 1)],
\lb{LEAG}
\eq
where the tensor ${K^{ab}}_{mn}$ is defined as
\bq 
{K^{ab}}_{mn} = c_1 g^{ab}g_{mn}+c_2\delta^{a}_{m} \delta^{b}_{n}
+c_3\delta^{a}_{n}\delta^{b}_{m}-c_4u^a u^b g_{mn},
\lb{Kab}
\eq
being the $c_i$ dimensionless coupling constants, and $\lambda$
a Lagrange multiplier enforcing the unit timelike constraint on the aether, and 
\bq
\delta^a_m \delta^b_n =g^{a\alpha}g_{\alpha m} g^{b\beta}g_{\beta n}.
\eq
The last term, $L_{\rm matter}$, is the matter Lagrangian, which depends on the metric tensor and the matter field.

In the weak-field, slow-motion limit EA theory reduces to Newtonian gravity with a value of  Newton's constant $G_{\rm N}$ related to the parameter $G$ in the action (\ref{action}) by  \cite{Garfinkle2007},
\bq
G = G_N\left(1-\frac{c_{14}}{2}\right).
\lb{Ge}
\eq
Here, the constant $c_{14}$ is defined as
\bq
c_{14}=c_1+c_4.
\lb{beta}
\eq

The field equations are obtained by extremizing the action with respect to independent variables of the system. The variation with respect to the Lagrange multiplier $\lambda$ imposes the condition that $u^a$ is a unit timelike vector, thus 
\bq
g_{ab}u^a u^b = -1,
\label{LagMul}
\eq
while the variation of the action with respect $u^a$, leads to \cite{Garfinkle2007}
\bq
\nabla_a J^a_b + c_4 u^b \nabla_b u_a \nabla_b u^a + \lambda u_b = 0,
\lb{lamu}
\eq
where,
\bq
J^a_m=K^{ab}_{mn} \nabla_b u^n.
\lb{Jdef}
\eq
The variation of the action with respect to the metric $g_{mn}$ gives the dynamical equations,
\bq
G^{Einstein}_{ab} = T^{aether}_{ab} +8 \pi G  T^{matter}_{ab},
\label{EA}
\eq
where
\bqn
G^{Einstein}_{ab} &=& R_{ab} - \frac{1}{2} g_{ab} R, \nb\\
T^{aether}_{ab}&=& \nabla_c [ J^c\;_{(a} u_{b)} + u^c J_{(ab)} - J_{(a} \;^c u_{b)}] \nb \\
& &- \frac{1}{2} g_{ab} J^c\;_d \nabla_c u^d+ \lambda u_a u_b  \nb \\
& & + c_1 [\nabla_a u_c \nabla_b u^c - \nabla^c u_a \nabla_c u_b] \nb \\
& & + c_4 u^c \nabla_c u_a u^c \nabla_c u_b, \nb \\
G^{EA}_{ab}&=&G^{Einstein}_{ab} - T^{aether}_{ab}, \nb\\
T^{matter}_{ab} &=&  \frac{- 2}{\sqrt{-g}} \frac{\delta \left( \sqrt{-g} L_{matter} \right)}{\delta g_{ab}}.
\label{fieldeqs}
\eqn

In a more general situation, the Lagrangian of GR is recovered if and only if the coupling constants are identically zero, e.g., $c_1=c_2=c_3=c_4=0$, 
considering the equations (\ref{Kab}) and (\ref{LagMul}).

\section{Cylindrical gravitational waves}

Let us start with the most general static cylindrical symmetric, given by
\bqn
ds^2 &=&  -e^{2 H(r, t)-2 \Psi(r, t)} dt^2 + e^{2 H(r, t)-2 \Psi(r, t)} dr^2 \nb \\
&& + e^{-2 \Psi(r, t)} r^2 d\phi^2 +e^{2 \Psi(r, t)} dz^2.
\lb{ds2}
\eqn
We assume that the indices of the Riemann e Einstein tensors as $(0,1,2,3)$ corresponding to the coordinates $(t,r,\phi,z)$, respectively. 
In accordance with equation (\ref{LagMul}), the aether field is assumed unitary, timelike, and constant, chosen as
\bq
u^a=(e^{\Psi(r, t)-H(r, t)},0,0,0).
\eq

	Assuming (\ref{ds2}), we compute the different terms in the field equations {
	(\ref{EA}), giving 
	\bqn
	&&(-2 H_{1} \Psi_{1} r-2 r H_{0} \Psi_{0}-2 H_{1}+3 \Psi_{0}^2 r +2 \Psi_{1} \nb\\ 
	&& +\Psi_{1}^2 r-2 r H_{11}+r H_{0}^2+2 \Psi_{11} r+r H_{1}^2) c_{13} \nb\\ 
	&& +(\Psi_{0}^2 r-2 r H_{0} \Psi_{0}+r H_{0}^2) c_{2} \nb\\
	&&+(2 r H_{11}-2 \Psi_{1}-2 \Psi_{11} r-\Psi_{1}^2 r+2 H_{1} \nb\\ 
	&& +2 H_{1} \Psi_{1} r-r H_{1}^2) c_{3}+ 
	(-2 H_{1}-2 H_{1} \Psi_{1} r \nb\\ 
	&& +r H_{1}^2 -2 r H_{11}+2 \Psi_{1}+\Psi_{1}^2 r+2 \Psi_{11} r) c_{4} \nb\\ 
	&& + 2 H_{1}-2 \Psi_{1}^2 r-2 \Psi_{0}^2 r =0,
	\lb{G00}
	\eqn
	\bqn
	&&(\Psi_{0}  H_{1} r-3 \Psi_{1} \Psi_{0} r-H_{1} H_{0} r+H_{0}+H_{0} \Psi_{1} r) c_{13}\nb\\ 
	&& + (H_{0} \Psi_{1} r-\Psi_{1} \Psi_{0} r-H_{1} H_{0} r+\Psi_{0}  H_{1} r) c_{2} \nb\\
	&& - H_{0}+2 \Psi_{1} \Psi_{0} r=0,
	\lb{G01}
	\eqn
	\bqn
	&&(H_{0} \Psi_{1} -\Psi_{1} \Psi_{0} -H_{1} H_{0} +\Psi_{0}  H_{1}) r c_{14}-\nb\\
	&&H_{0}+2 \Psi_{1} \Psi_{0} r=0,
	\lb{G10}
	\eqn
	\bqn
	&&(-2 r H_{00}-2 r H_{0} \Psi_{0}+r H_{0}^2+3 \Psi_{0}^2 r+r H_{1}^2\nb\\
	&&+\Psi_{1}^2 r+2 r \Psi_{00}-2 H_{1} \Psi_{1} r) c_{13}\nb\\
	&&+(r H_{0}^2+2 r \Psi_{00}-2 r H_{0} \Psi_{0}- 2 r H_{00}+\Psi_{0}^2 r) c_{2}\nb\\
	&&+(2 H_{1} \Psi_{1} r-r H_{1}^2-\Psi_{1}^2 r) c_{3}\nb\\
	&&+(-2 H_{1} \Psi_{1} r+r H_{1}^2+\Psi_{1}^2 r) c_{4}\nb\\
	&&+2 H_{1}-2 \Psi_{1}^2 r-2 \Psi_{0}^2 r =0,
	\lb{G11}
	\eqn
	\bqn
	&&(3 \Psi_{0}^2+2 H_{1} \Psi_{1}-H_{1}^2-\Psi_{1}^2+2 \Psi_{00}\nb\\
	&&+H_{0}^2-2 \Psi_{0} H_{0}) c_{13}+\nb\\
	&&(H_{0}^2+2 \Psi_{00}-2 \Psi_{0} H_{0}-2 H_{00}+\Psi_{0}^2) c_{2}\nb\\
	&&+(-2 H_{1} \Psi_{1}+H_{1}^2+\Psi_{1}^2) c_{3}\nb\\
	&&+(2 H_{1} \Psi_{1}-H_{1}^2-\Psi_{1}^2) c_{4}\nb\\
	&&+2 \Psi_{1}^2-2 H_{00}+2 H_{11}-2 \Psi_{0}^2 =0,
	\lb{G22}
	\eqn
	\bqn
	&&(-r H_{0}^2 + 2 r H_{0} \Psi_{0}+\Psi_{1}^2 r+2 r \Psi_{00} -3 \Psi_{0}^2 r\nb\\
	&&-2 H_{1} \Psi_{1} r+r H_{1}^2) c_{13} +(-r H_{0}^2+2 r H_{0} \Psi_{0}\nb\\
	&&-\Psi_{0}^2 r+2 r H_{00}-2 r \Psi_{00}) c_{2} +(2 H_{1} \Psi_{1} r \nb\\
	&& -r H_{1}^2 -\Psi_{1}^2 r) c_{3} +(-2 H_{1} \Psi_{1} r+r H_{1}^2\nb\\
	&&+\Psi_{1}^2 r) c_{4}+4 \Psi_{1}-2 \Psi_{1}^2 r+2 \Psi_{0}^2 r-2 r H_{11}\nb\\
	&&+4 \Psi_{11} r-4 r \Psi_{00}+2 r H_{00} =0,
	\lb{G33}
	\eqn
	where the subscript zero denotes the derivative with respect to the coordinate $t$, the subscript one denotes the derivative in relation to the coordinate $r$} and $G^{aether}_{\mu\nu}=G^{Einstein}_{\mu\nu}-T^{aether}_{\mu\nu}$. We assume also $c_{13}=c_1+c_3$.

\section{Gravitational Wave Equation}

Obtaining $H_0$ from the  
equation (\ref{G10}), we get
	\bqn
	H_{0} =  \frac{\Psi_{0} r [ 2 \Psi_{1}  +c_{14}(H_{1}-\Psi_{1})]}{1 -c_{14}  r(H_{1}-\Psi_{1})},
	\lb{H_0}
	\eqn
	and $H_{00}$ from the 
	equation (\ref{G33}), we have
	\bqn
	H_{00} &=& \frac{1}{2r (1+c_{2})} \left(-4 \Psi_{1}+2 \Psi_{1}^2 r -2 \Psi_{0}^2 r+2 c_{14} H_{1} \Psi_{1} r \right.\nb\\
	&&\left. -2 c_{2} r H_{0} \Psi_{0}+c_{13} r H_{0}^2+ 2 r H_{11}+4 r \Psi_{00}\right.\nb\\
	&&\left. -4 \Psi_{11} r -c_{14} r H_{1}^2+c_{2} r H_{0}^2-c_{14} \Psi_{1}^2 r+ c_{2} \Psi_{0}^2 r\right.\nb\\
	&&\left.+2 c_{2} r \Psi_{00} -2 c_{13} r H_{0} \Psi_{0}+ 3 c_{13} \Psi_{0}^2 r
	-2 c_{13} r \Psi_{00}\right),
	\lb{H_00}
	\eqn
	$c_2 \neq -1$
	and $H_1$ from the
	equation (\ref{G01}), we obtain
	\bqn
	H_{1}&=& \frac{1}{r (c_{13}+c_{2})(-\Psi_0+H_0)}\left(-H_{0}\right.\nb\\
	&&\left. +2 \Psi_{1} \Psi_{0} r+ c_{13} H_{0} \Psi_{1} r-3 c_{13} \Psi_{1} \Psi_{0} r+c_{2} H_{0} \Psi_{1} r\right.\nb\\
	&&\left.-c_{2} \Psi_{1} \Psi_{0} r+c_{13} H_{0}\right).
	\lb{H_1}
	\eqn
	
	Here, we have checked the possible existence of a bifurcation of solutions.
We have two possibilities from equation (\ref{H_1}): (i) $c_2=c_{13}=0$ or (ii) $c_2=-c_{13}$. In general we can assume that $H_0 \neq \Psi_0$.  In the case (i) we obtain
from the equation (\ref{G01}) and from equation (\ref{G10}) the quantity $H_0$ that
must be identical to each other in order to have system consistency, i.e., $\Psi_0 r (-1+2\Psi_1 r)(-c_4+c_3)(H_1-\Psi_1)=0$. Again, assuming in general that $H_1 \neq \Psi_1$ and 
substituting the condition $c_3=c_4$ into equations
(\ref{G00})-(\ref{G33}) we obtain exactly the GR field equations. In the case (ii)
we obtain again from the equation (\ref{G01}) and from equation (\ref{G10}) the quantity $H_0$ that
must be identical to each other in order to have system consistency, i.e., $\Psi_0 r (-1+2\Psi_1 r)(c_{14})(H_1-\Psi_1)=0$.  Again, assuming in general that $H_1 \neq \Psi_1$ and 
substituting the condition $c_{14}=0$ into equations
(\ref{G00})-(\ref{G33}) we obtain exactly the EA field equations. These two cases
suggest no GR limit for EA wave equations.

We can obtain $H_{11}$ from the equation (\ref{G22}), we compute
	\bqn
	H_{11}&=& \frac{1}{2} c_{14} H_{1}^2+H_{00}-\Psi_{1}^2+\Psi_{0}^2-\frac{1}{2} c_{2} H_{0}^2+\nb\\
	&&\frac{1}{2} \Psi_{1}^2 c_{14}-\frac{1}{2} c_{2} \Psi_{0}^2+c_{2} H_{00}-c_{2} \Psi_{00}-\Psi_{1} H_{1} c_{14}\nb\\
	&&+c_{2} \Psi_{0} H_{0}-\frac{3}{2} c_{13} \Psi_{0}^2-\frac{1}{2} c_{13} H_{0}^2-c_{13} \Psi_{00}+c_{13} \Psi_{0} H_{0}.
	\lb{H_11}
	\eqn

Substituting equations (\ref{H_0})-(\ref{H_11}) into (\ref{G22}) we obtain
the wave equation \cite{Jacobson2004}
\bqn
{\Psi_{00}}= \frac{1}{ 1-{c_{13}}}\left( \Psi_{11}+ \frac{\Psi_{1}}{r} \right).
\lb{wave}
\eqn

Obtaining $\Psi_1$ from equation (\ref{wave}) we get
\bqn
\Psi_{1}= -r (\Psi_{11}-\Psi_{00}+c_{13} \Psi_{00}).
\lb{Psi_1}
\eqn

Substituting equations (\ref{H_1}) and (\ref{Psi_1}) into the
equation (\ref{G01}), we obtain an identity.

Substituting equations (\ref{H_0}) and (\ref{Psi_1}) into the 
equation (\ref{G10}) we also obtain an identity.

Substituting equations (\ref{H_0})-(\ref{H_11}) and (\ref{Psi_1}) into the  
equations (\ref{G22}) and (\ref{G33}), we obtain identities.

The components of the field equations that we have not used yet are 
equations (\ref{G00}) and (\ref{G11}).
From equations (\ref{G00}) and (\ref{G11}) we get that
\bqn
&&(-2 H_1+2 \Psi_{1}+2 \Psi_{11} r-2 r H_{11}-2 r \Psi_{00}+2 r H_{00}) c_{13}\nb\\
&&+(2 r H_{00}-2 r \Psi_{00}) c_{2}+(2 H_1-2 \Psi_{11} r+2 r H_{11}-2 \Psi_{1}) c_{3}\nb\\
&&+(2 \Psi_{1}+2 \Psi_{11} r-2 r H_{11}-2 H_1) c_{4}=0.
\lb{bound}
\eqn
Thus, besides the wave equation (\ref{wave}), we have an additional boundary equation (\ref{bound}) that has to be satisfied in order for gravitational waves to exist in EA theory. Thus, we can calculate also the metric function $H(t,r)$. Notice that in GR where $c_2 = 0$, $c_3 = 0$, $c_4 = 0$ and $c_{13} = 0$, we do not have this additional equation. Once we have calculated $\Psi(t,r)$ from the wave equation, we can return to one of the field equations' components and solve it to obtain $H(t,r)$. Assuming $c_2=0$, $c_3=0$, $c_4=0$, and $c_{13}=0$ in all the calculated results, we obtain exactly the GR results.

\section{Solution of the Gravitational Wave Equation}

From the equation (\ref{wave}), we can identify that the velocity of the 
wave is
\bq
c_{ea}={\frac {1}{\sqrt {1-{c_{13}}}}},
\eq
and the wave number is given by 
\bq
{k}={\mathrm w}\,\sqrt {1-{c_{13}}},
\lb{k}
\eq
where ${\mathrm w}$ is the frequency of the gravitational wave.  Notice that in EA, the velocity of the  wave can be infinity when $c_{13} \rightarrow 1$, while in GR the velocity of the  wave is the speed of the light. 
{We also must have $c_{13}<1$ in order 	to have a wave.}

Assuming a stationary wave, we have the monochromatic solution as
\bq
\Psi {(r,t)}=f \left( r \right) \cos \left( {\mathrm w}\,t \right),
\lb{Psir}
\eq
and substituting this equation into equation (\ref{wave}) we get
\bq
f \left( r \right) ={C_1}\,J_0\left(k\,r\right)+{C_2}\,Y_0\left(k\,r\right),
\lb{fr}
\eq
where $C_1$, $C_2$ are arbitrary constants of integration and
where $J_n(r)$ and $Y_n(r)$ are the Bessel of the first and second kind
of order $n$, respectively.

Substituting equation (\ref{Psir}) and (\ref{fr}) into (\ref{bound})
we obtain that for a monochromatic wave 
\bqn
H \left( r,t \right)&=&{F_1} \left( r \right) {F_2} \left( t
\right) + \left[ {C_6}\,J_0\left({C_5}\,r\right)+
{C_7}\,Y_0\left({C_5}\,r\right) \right]  \times\nb\\
&&\left[ {C_3}\,{e}^{g(r,t)}+{C_4}\,{{e}^{-g(r,t)}} \right],
\eqn
where
\bq
{F_1(r)}={C_8}\,
J_0\left(\sqrt {C_{12}}\,r\right)+{C_9}\,Y_0\left(\sqrt {C_{12}}\,r\right),
\eq
	\bqn
	{F_2(t)}&=&{C_{10}}\,\cos \left( {\frac {\sqrt {{c_{14}}}\,\sqrt {C_{12}}}{\sqrt {{c_{2}}+{c_{13}}}}}\,\,t \right) +{C_{11}}\,\sin \left( {\frac {\sqrt {c_{14}}\,\sqrt {C_{12}}}{\sqrt {{c_{2}}+{c_{13}}}}}\,\,t \right),
	\lb{F2t}
	\eqn
and
\bqn
g(r,t)&=&\sqrt {c_{14}}\times\nb\\
&&\left\{ \left[ -{C_6} \left( J_0\left({C_5}\,r\right)-
\frac {J_1\left({C_5}\,r\right)}{{C_5}\,r}\right) - \right.\right.\nb\\
&&\left.\left. {C_7} \left( 
Y_0\left({C_5}\,r\right)-\frac {
	Y_1\left({C_5}\,r\right)}{{C_5}\,r} \right) \right] {{C_5}}^{2}\,r-\right.\nb\\
&&\left.{C_5}\,{C_6}\,\,
J_1\left({C_5}\,r\right)-{C_5}\,{C_7}\,\,
Y_1\left({C_5}\,r\right) \right. \Bigg\}^{\frac{1}{2}} \times\nb\\
&&{\frac {1}{\sqrt {r
}}}\,\,{\frac {1}{\sqrt {{C_6}\,
			J_0\left({C_5}\,r\right)+{C_7}\,
			Y_0\left({C_5}\,r\right)}}}\,\,{\frac {1}{\sqrt {{c_{2}}+{c_{13}}}}}\,\,t,
\eqn
where $C_3$, $C_4$, $C_5$, $C_6$, $C_7$, $C_8$, $C_9$, $C_{10}$, $C_{11}$
and $C_{12}$ are arbitrary constants of integration.

Since the Bessel function of the second kind is singular at $r=0$, we will
assume that  $C_2$, $C_7$ and $C_9$ are null, thus we have
\bq
f \left( r \right) ={C_1}\,J_0\left(k\,r\right),
\eq
\bq
{F_1(r)}={C_8}\,J_0\left(\sqrt {C_{12}}\,r\right),
\eq
\bqn
H(r,t)&=&{F_1} \left( r \right) {F_2} \left( t \right) \nb \\
&&+{C_6}\,
{J_0\left({C_5}\,r\right)} \left[ {C_3}\,{{e}^{g(r,t)}}+{C_4}\,{{e}^{-g(r,t)}} \right].
\eqn

Simplifying $g(r,t)$ we get
	\bqn
	g(r,t)\equiv g(t)= i\, \frac {\sqrt { c_{14}}\,\,C_5}
	{\sqrt { c_{13}+ c_{2}}}\,t,
	\eqn
where $i=\sqrt{-1}$ is the imaginary constant.
However, the equation (\ref{F2t}) remains unchanged.

Thus, redefining the constants $C_3$ and $C_4$, we can have
	\bqn
	H(r,t)&=&{C_8}\,{J_0\left(\sqrt {C_{12}}\,r\right)} \times\nb\\
	&&\left[ {C_{10}}\,\cos \left( {\frac {\sqrt {c_{14}}\,\sqrt {C_{12}}}{\sqrt {{c_{13}}+{c_2}}}}\,\,t \right) +{C_{11}}\,\sin \left( {\frac {\sqrt {c_{14}}\,\sqrt {C_{12}}}{\sqrt {{c_{13}}+{c_2}}}}\,\,t \right)  \right] +\nb\\
	&&{C_6}\,{J_0\left({C_5}\,r\right)} \times\nb\\
	&&\left[ {C_3}\,\cos
	\left( {\frac {\sqrt {c_{14}}\,{C_5}}{\sqrt {{c_{13}}+{c_2}}}}\,\,t \right) +{C_4}\,\sin \left( {\frac {
			\sqrt {c_{14}}\,{C_5}}{\sqrt {{c_{13}}+{c_2}}}}\,\,t \right) \right].
	\eqn
	Substituting these wave solutions into the equations (\ref{G00})-(\ref{G33}) 
	we must have
\bq
{c_2}=-{c_3}-{{c_{13}}}^{2}-{c_{13}}\,{c_4}+{c_4}+{c_{13}}\,{c_3}.
\lb{c2a}
\eq

Let us suppose that $\sqrt{C_{12}}=C_5=\omega$, $C_8=C_6$, $C_{11}=C_{4}$, $C_{10}=C_3$,
for the sake of simplicity, and using equation (\ref{k}), we can rewrite the metric functions as
\bq
\Psi {(r,t)}={C_1}\, \cos \left( {\mathrm w}\,t \right)\,J_0\left(\sqrt {1-{c_{13}}}\,{\mathrm w}\,r\right),
\lb{Psirt}
\eq

	\bqn
	H(r,t)&=&2{C_8}\,J_0\left(\omega \,r\right)\left[ {C_{3}}\,\cos \left( \frac {\sqrt {c_{14}}}{\sqrt {{c_{13}}+{c_2}}}\,\omega\,t \right) +{C_{4}}\,\sin \left( \frac {\sqrt {c_{14}}}{\sqrt {{c_{13}}+{c_2}}}\, \omega \,t \right)\right],\nb\\
	\eqn

Let us suppose that $C_8=1/2$, $C_3=1$, $C_{4}=0$ for the sake of simplicity.
Thus, we can rewrite the metric functions as

\bqn
H(r,t)&=&\cos \left( \frac {\sqrt {c_{14}}}{\sqrt {{c_{13}}+{c_2}}}\,\omega\,\,t \right) J_0\left(\omega\,r\right).
\lb{Hrt}
\eqn

Comparing equations (\ref{Psirt}) and (\ref{Hrt}), we can see that they have the 
same functional dependence in the coordinates $r$ and $t$, 
but with different frequencies in both coordinates, ${\mathrm w}$ and $\omega$. 
Section 6 will show that this temporal behavior
is different in GR.

\section{Gravitational Wave Pulse with Width}

Let us now calculate the superposition of waves of the type of (\ref{Psirt}) 
and (\ref{Hrt}) with amplitude factor as $\exp(-\beta\, {\mathrm w})$.  
Here, we will follow the same procedure used in reference \cite{Weber1957}.
Thus, we have 
\bq
\Psi=\int _{0}^{\infty }\!{{\rm e}^{-{\beta}\,{\mathrm w}}}\cos \left( {\mathrm w}\,t \right) 
J_0\left({\mathrm w}\sqrt {1-{c_{13}}}r \right){d{\mathrm w}}.
\eq

From the reference \cite{Gradshteyn2007} we have that
\bqn
&&\int _{0}^{\infty }\!{{\rm e}^{-{\beta}\,{\mathrm w}}}\cos \left( {\alpha}\,{\mathrm w}
\right) J_0\left({\gamma}\,{\mathrm w}\right){d{\mathrm w}}=\nb\\
&&\frac{\sqrt {2}}{2}\,{\frac {\sqrt {\sqrt {{{\beta}}^{4}+2\,{{\beta}}^{2}{{\gamma}}^
				{2}+2\,{{\alpha}}^{2}{{\beta}}^{2}+{{\gamma}}^{4}-2\,{{\gamma}}^{2}{{
						\alpha}}^{2}+{{\alpha}}^{4}}+{{\beta}}^{2}+{{\gamma}}^{2}-{{\alpha}}^
			{2}}}{\sqrt {{{\beta}}^{4}+2\,{{\beta}}^{2}{{\gamma}}^{2}+2
			\,{{\alpha}}^{2}{{\beta}}^{2}+{{\gamma}}^{4}-2\,{{\gamma}}^{2}{{
					\alpha}}^{2}+{{\alpha}}^{4}}}},
\lb{intpulse}
\eqn
\bqn
&&\Psi(r,t)= \nb\\
&&\frac{\sqrt {2}}{2}{\frac {\sqrt {\sqrt {{{\beta}}^{4}+2\,{{\beta}}^{2}{r}^{2}+2\,
				{t}^{2}{{\beta}}^{2}+{r}^{4}-2\,{r}^{2}{t}^{2}+{t}^{4}}+{{\beta}}^{2
			}+{r}^{2}-{t}^{2}}}{\sqrt {{{\beta}}^{4}+2\,{{\beta}}^{2}{r
			}^{2}+2\,{t}^{2}{{\beta}}^{2}+{r}^{4}-2\,{r}^{2}{t}^{2}+{t}^{4}}}},
\eqn
where $\alpha=t$ and $\gamma=\sqrt{1-c_{13}}\, r$.

Let us make the following coordinate transformation
\bq
r = -t+\beta\, x
\eq

assuming that $\beta=1$, $t=0$ or $t=2$ or $t=6$ or $t=10$ and $c_{13}=0$ or $c_{13}=\pm 1/2$, we can write that
\bqn
\Psi(x,t=0,c_{13}=1/2)&=&\,{\frac {\sqrt {2}}{\sqrt {2+{x}^{2}}}},
\eqn
\bqn
&&\Psi(x,t=2,c_{13}=1/2)=\nb\\
&&\frac{1}{2}\,{\frac {\sqrt {2\,\sqrt {100-12\, \left( -2+x \right) ^{2}+
 \left( -2+x \right) ^{4}}-12+2\, \left( -2+x \right) ^{2}}\sqrt {2}}{
\sqrt {100-12\, \left( -2+x \right) ^{2}+ \left( -2+x \right) ^{4}}}},
\eqn
\bqn
&&\Psi(x,t=6,c_{13}=1/2)=\nb\\
&&\frac{1}{2}\,{\frac {\sqrt {2\,\sqrt {5476-140\, \left( -6+x \right) ^{2}+
 \left( -6+x \right) ^{4}}-140+2\, \left( -6+x \right) ^{2}}\sqrt {2}}
{\sqrt {5476-140\, \left( -6+x \right) ^{2}+ \left( -6+x \right) ^{4}}
}},
\eqn
\bqn
&&\Psi(x,t=10,c_{13}=1/2)=\nb\\
&&\frac{1}{2}\,{\frac {\sqrt {2\,\sqrt {40804-396\, \left( -10+x \right) ^{2}+
 \left( -10+x \right) ^{4}}-396+2\, \left( -10+x \right) ^{2}}\sqrt {2
}}{\sqrt {40804-396\, \left( -10+x \right) ^{2}+ \left( -10+x \right) 
^{4}}}},\nb\\
\eqn
\bqn
\Psi(x,t=0,c_{13}=0)&=&{\frac {1}{\sqrt {1+{x}^{2}}}},
\eqn
\bqn
&&\Psi(x,t=2,c_{13}=0)=\nb\\
&&\frac{1}{2}\,{\frac {\sqrt {\sqrt {25-6\, \left( -2+x \right) ^{2}+ \left( -2+
x \right) ^{4}}-3+ \left( -2+x \right) ^{2}}\sqrt {2}}{\sqrt {25-6\,
 \left( -2+x \right) ^{2}+ \left( -2+x \right) ^{4}}}},
\eqn
\bqn
&&\Psi(x,t=6,c_{13}=0)=\nb\\
&&\frac{1}{2}\,{\frac {\sqrt {\sqrt {1369-70\, \left( -6+x \right) ^{2}+ \left( 
-6+x \right) ^{4}}-35+ \left( -6+x \right) ^{2}}\sqrt {2}}{\sqrt {1369
-70\, \left( -6+x \right) ^{2}+ \left( -6+x \right) ^{4}}}},
\eqn
\bqn
&&\Psi(x,t=10,c_{13}=0)=\nb\\
&&\frac{1}{2}\,{\frac {\sqrt {\sqrt {10201-198\, \left( -10+x \right) ^{2}+
 \left( -10+x \right) ^{4}}-99+ \left( -10+x \right) ^{2}}\sqrt {2}}{
\sqrt {10201-198\, \left( -10+x \right) ^{2}+ \left( -10+x \right) ^{4
}}}}.
\eqn
\bqn
\Psi(x,t=0,c_{13}=-1/2)&=&{\frac {\sqrt {2}}{\sqrt {2+3\,{x}^{2}}}},
\eqn
\bqn
&&\Psi(x,t=2,c_{13}=-1/2)=\nb\\
&&\frac{1}{2}\,{\frac {\sqrt {2\,\sqrt {100-36\, \left( -2+x \right) ^{2}+9\,
 \left( -2+x \right) ^{4}}-12+6\, \left( -2+x \right) ^{2}}\sqrt {2}}{
\sqrt {100-36\, \left( -2+x \right) ^{2}+9\, \left( -2+x \right) ^{4}}
}},
\eqn
\bqn
&&\Psi(x,t=6,c_{13}=-1/2)=\nb\\
&&\frac{1}{2}\,{\frac {\sqrt {2\,\sqrt {5476-420\, \left( -6+x \right) ^{2}+9\,
 \left( -6+x \right) ^{4}}-140+6\, \left( -6+x \right) ^{2}}\sqrt {2}}
{\sqrt {5476-420\, \left( -6+x \right) ^{2}+9\, \left( -6+x \right) ^{
4}}}},\nb\\
\eqn
\bqn
&&\Psi(x,t=10,c_{13}=-1/2)=\nb\\
&&\frac{1}{2}\,{\frac {\sqrt {2\,\sqrt {40804-1188\, \left( -10+x \right) ^{2}+9
\, \left( -10+x \right) ^{4}}-396+6\, \left( -10+x \right) ^{2}}\sqrt 
{2}}{\sqrt {40804-1188\, \left( -10+x \right) ^{2}+9\, \left( -10+x
 \right) ^{4}}}}.\nb\\
\eqn
See these examples in Figures \ref{fig2} to \ref{fig7}, using the parameters
of the Table \ref{table1}.

\begin{table*}

\centering
\begin{minipage}{100 mm}
\caption{Summary of the Aether Parameters}
\label{table1}
\begin{tabular}{@{}|c|c|c|c|}
\hline
Case & $c_{13}$ & $c_{14}$ & $c_{2}$ \\
\hline
$A$  & -1/2  & 2 & 7/2 \\
\hline
$B$ & 0  & 1/2 & 1/2 \\
\hline
$C$ & 1/2 & 3/2  & 1/4 \\
\hline
\end{tabular}
\medskip\\
These parameters must obey the equation (\ref{c2a})
and the solar system tests, thus, we must also have $0 \le c_{14} < 2$, $c_{13}<1$ and $2+c_{13}+2c_2>0$ \cite{Foster:2005dk}.
\end{minipage}

\end{table*}

\begin{figure}[!ht]
	\centering	
	\includegraphics[width=7cm]{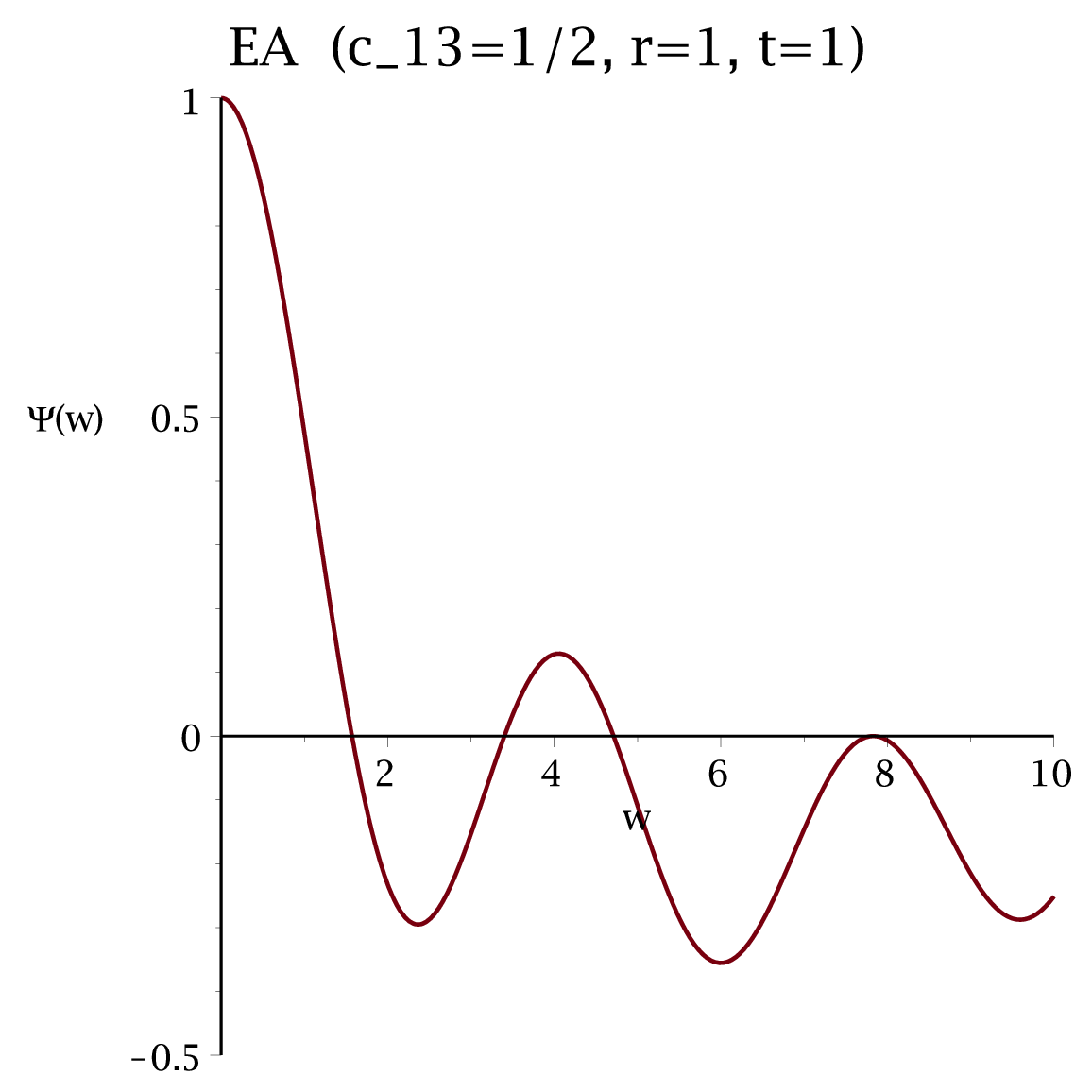}
	\includegraphics[width=7cm]{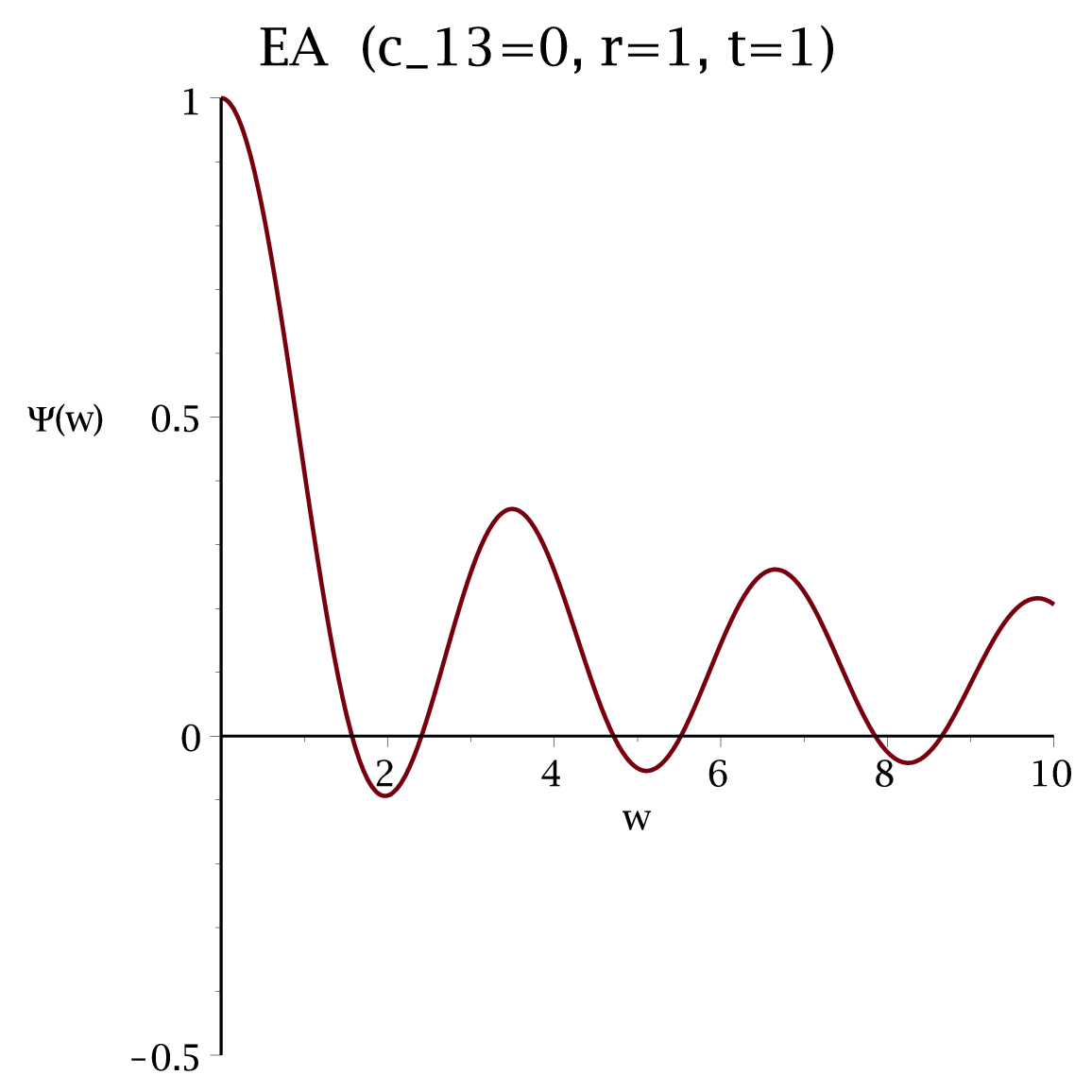}
	\includegraphics[width=7cm]{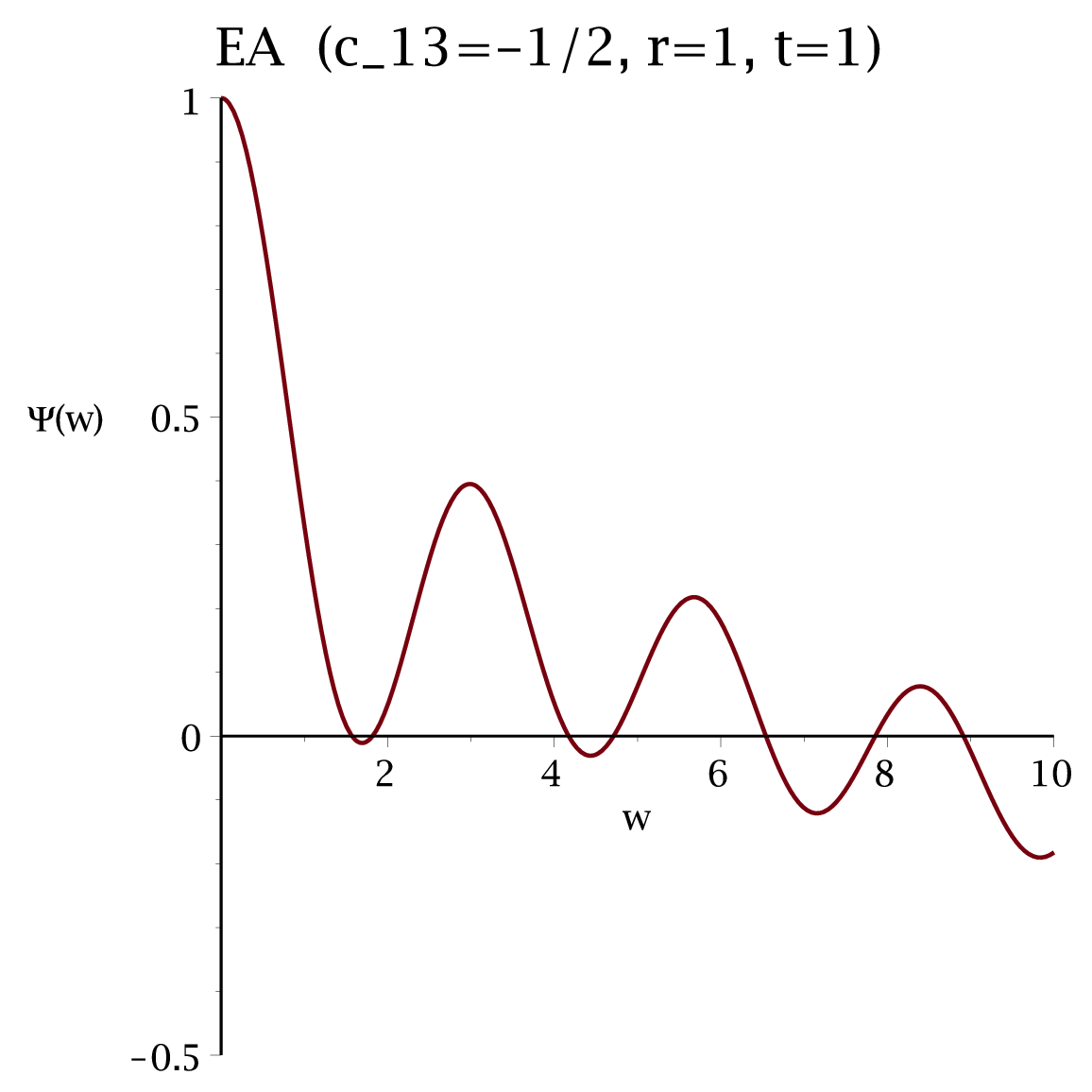}
	\includegraphics[width=7cm]{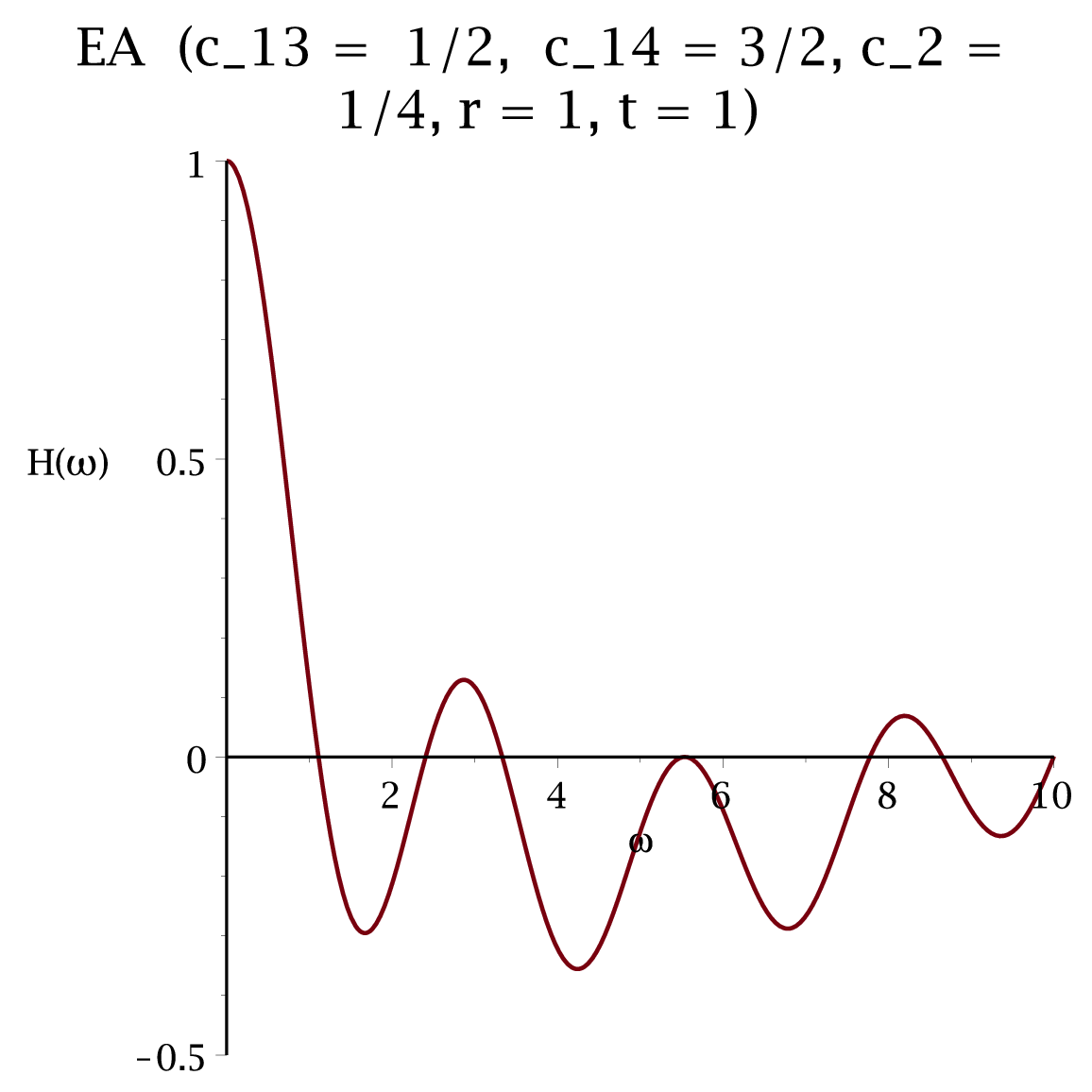}
	\includegraphics[width=7cm]{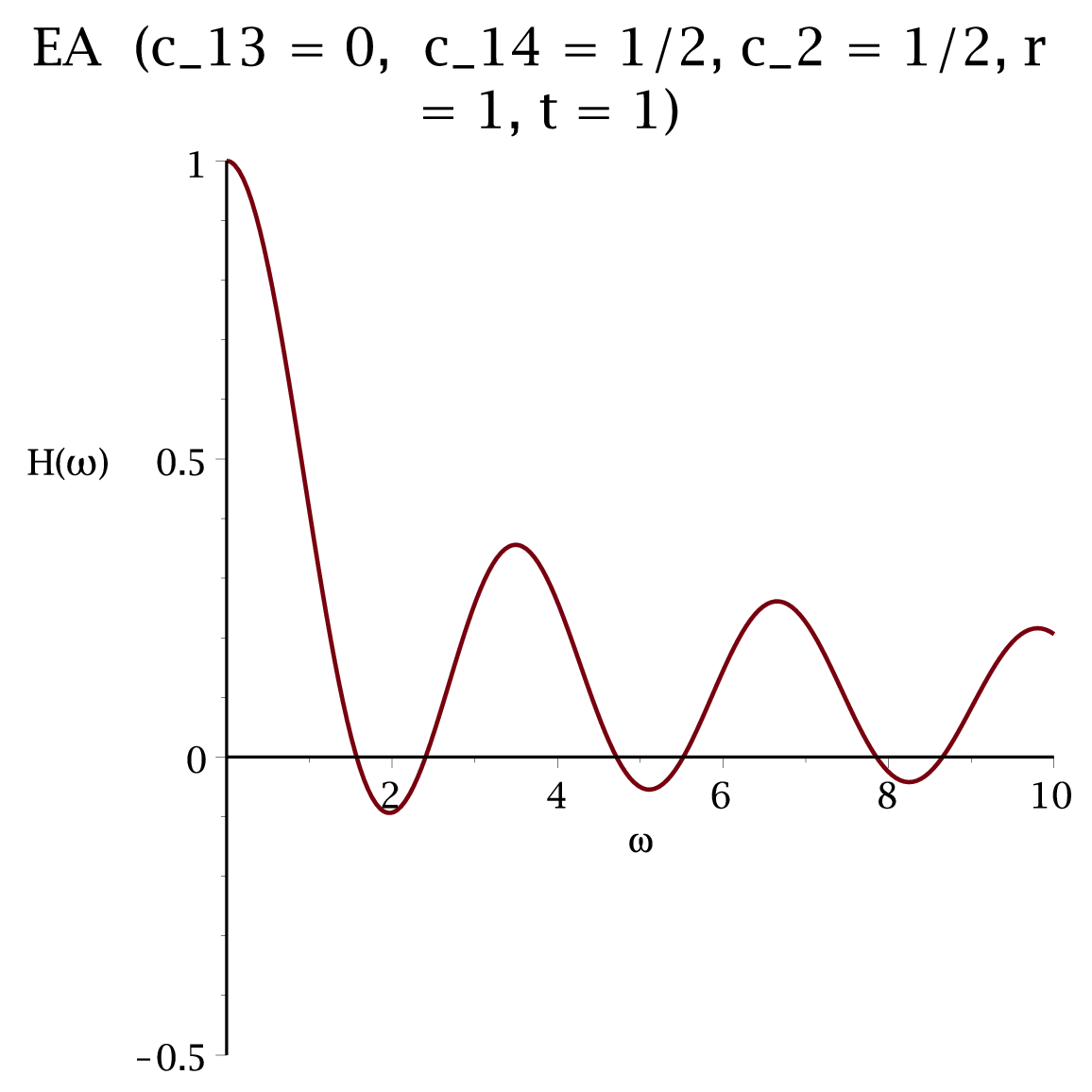}
	\includegraphics[width=7cm]{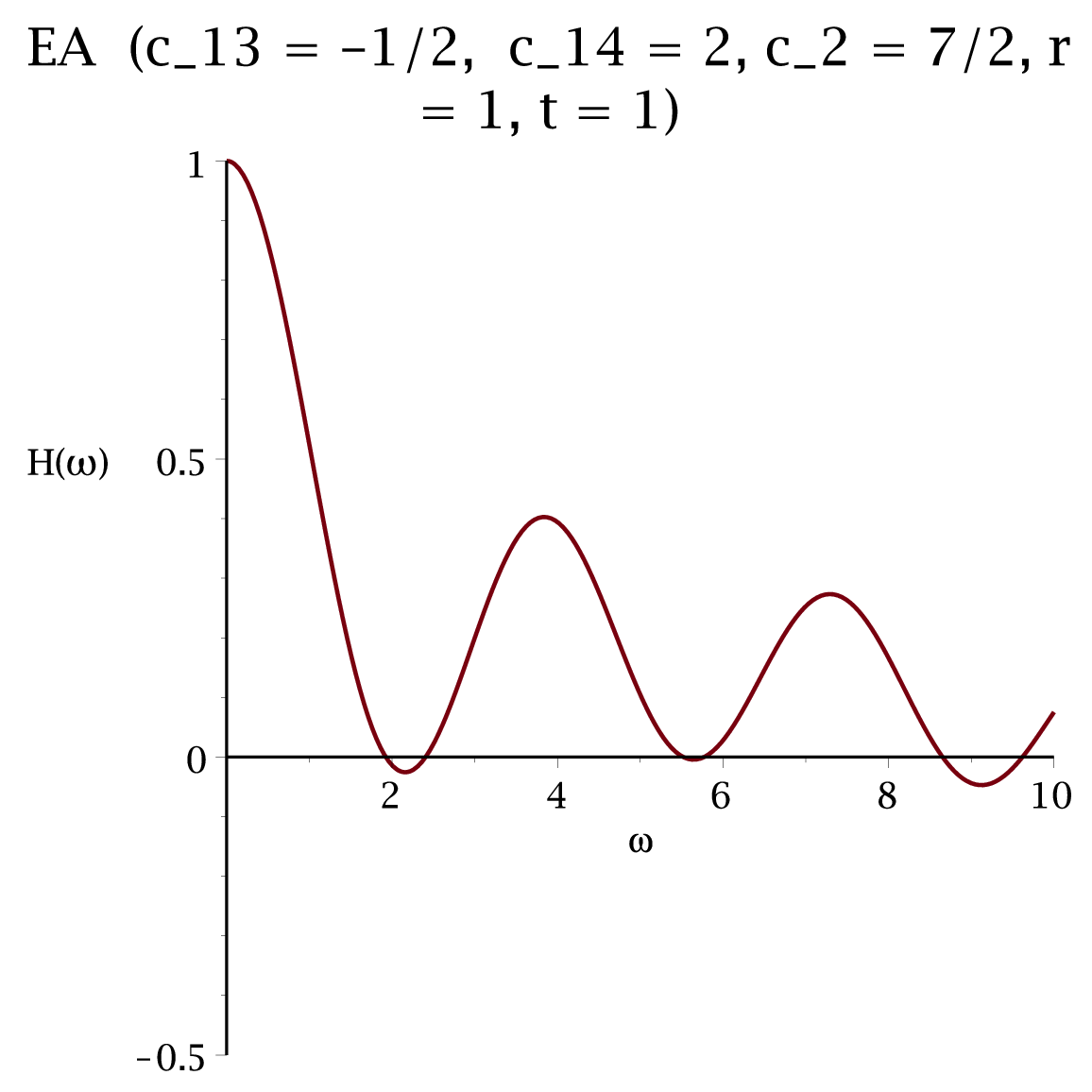}
	\caption{Wave components  $\Psi({\mathrm w})$ and $H(\omega)$ using $r=1$, $t=1$ and the parameters of Table \ref{table1} in equations (\ref{Psirt}) and (\ref{Hrt}).}
	\label{fig1}
\end{figure}

\begin{figure}[!ht]
	\centering	
	\includegraphics[width=7cm]{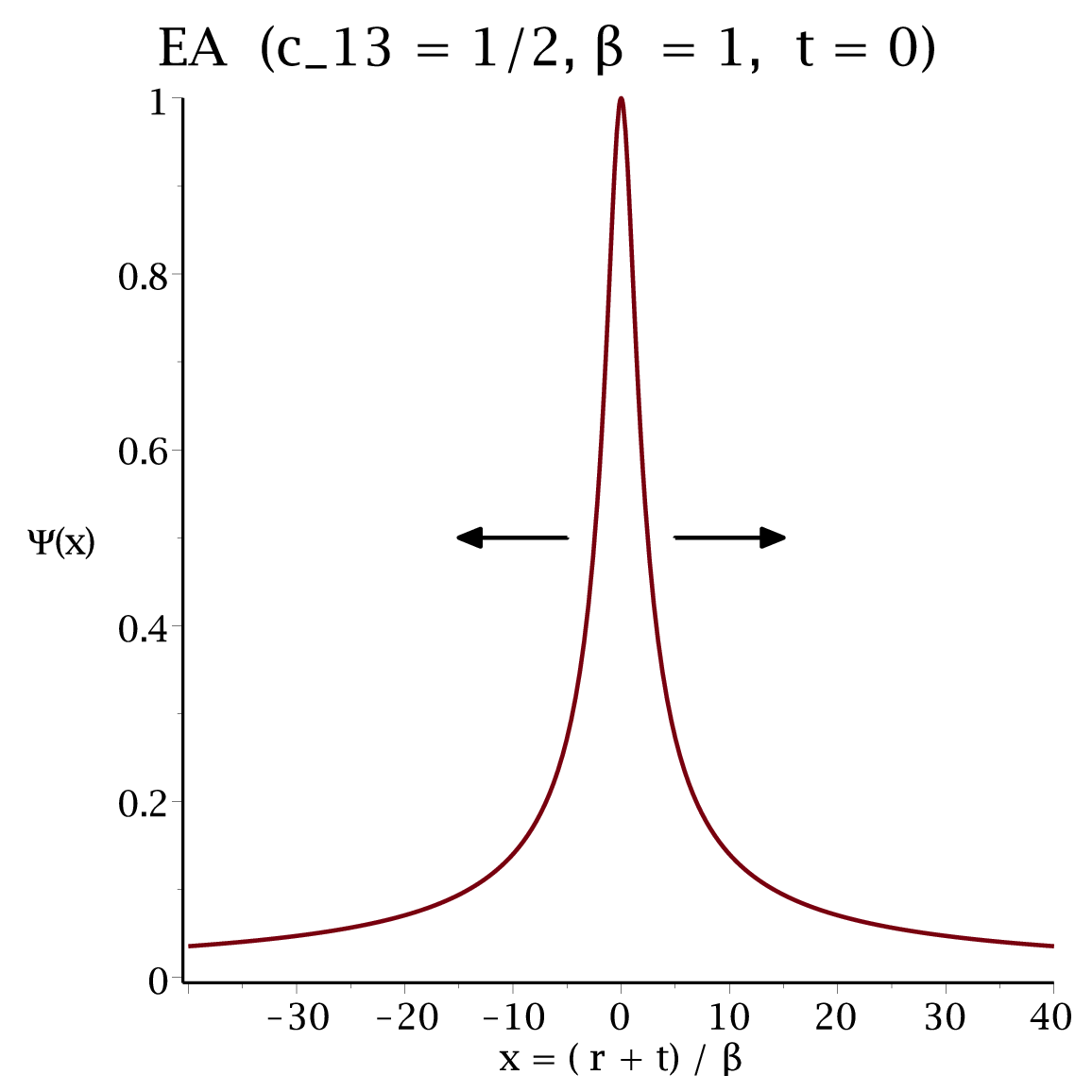}
	\includegraphics[width=7cm]{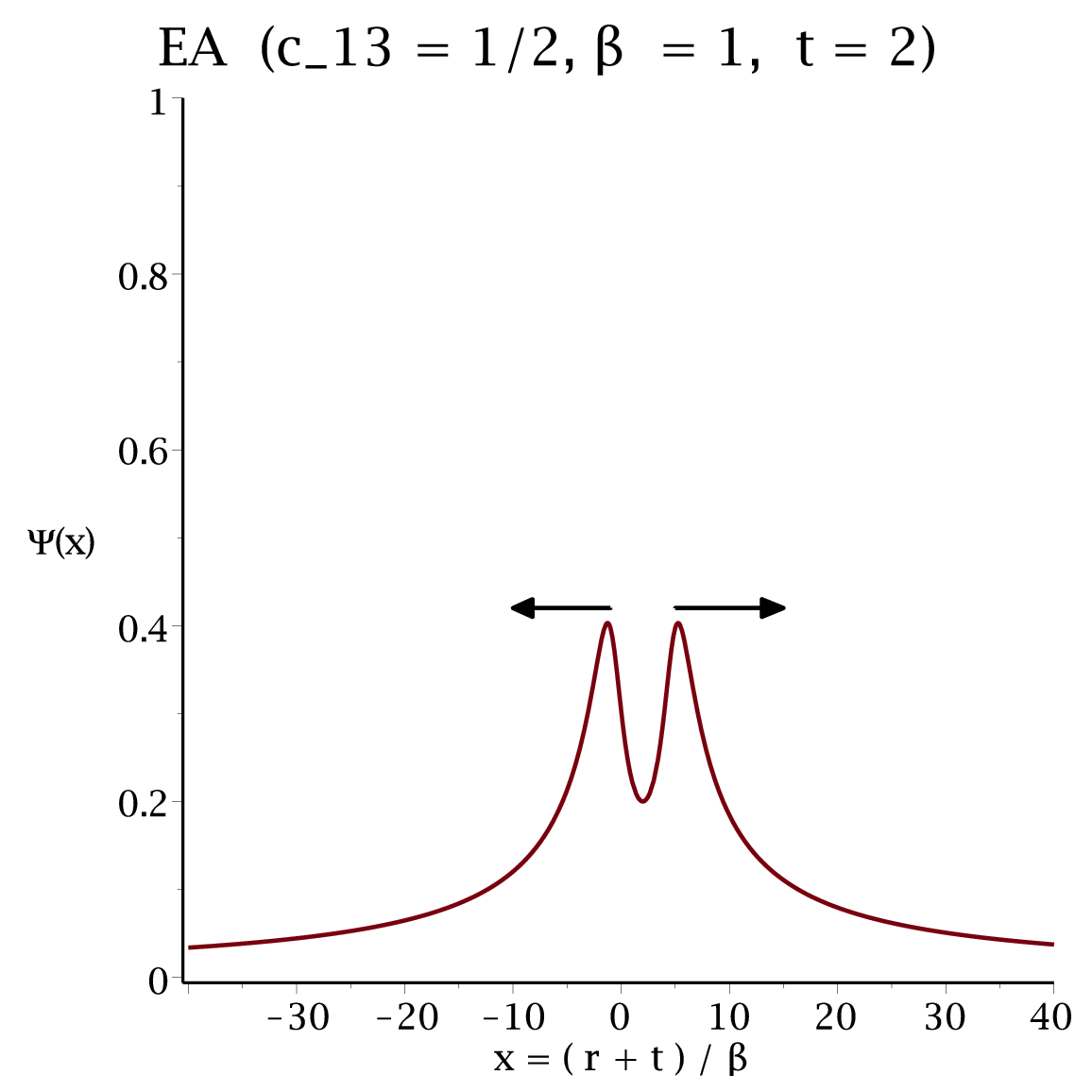}
	\includegraphics[width=7cm]{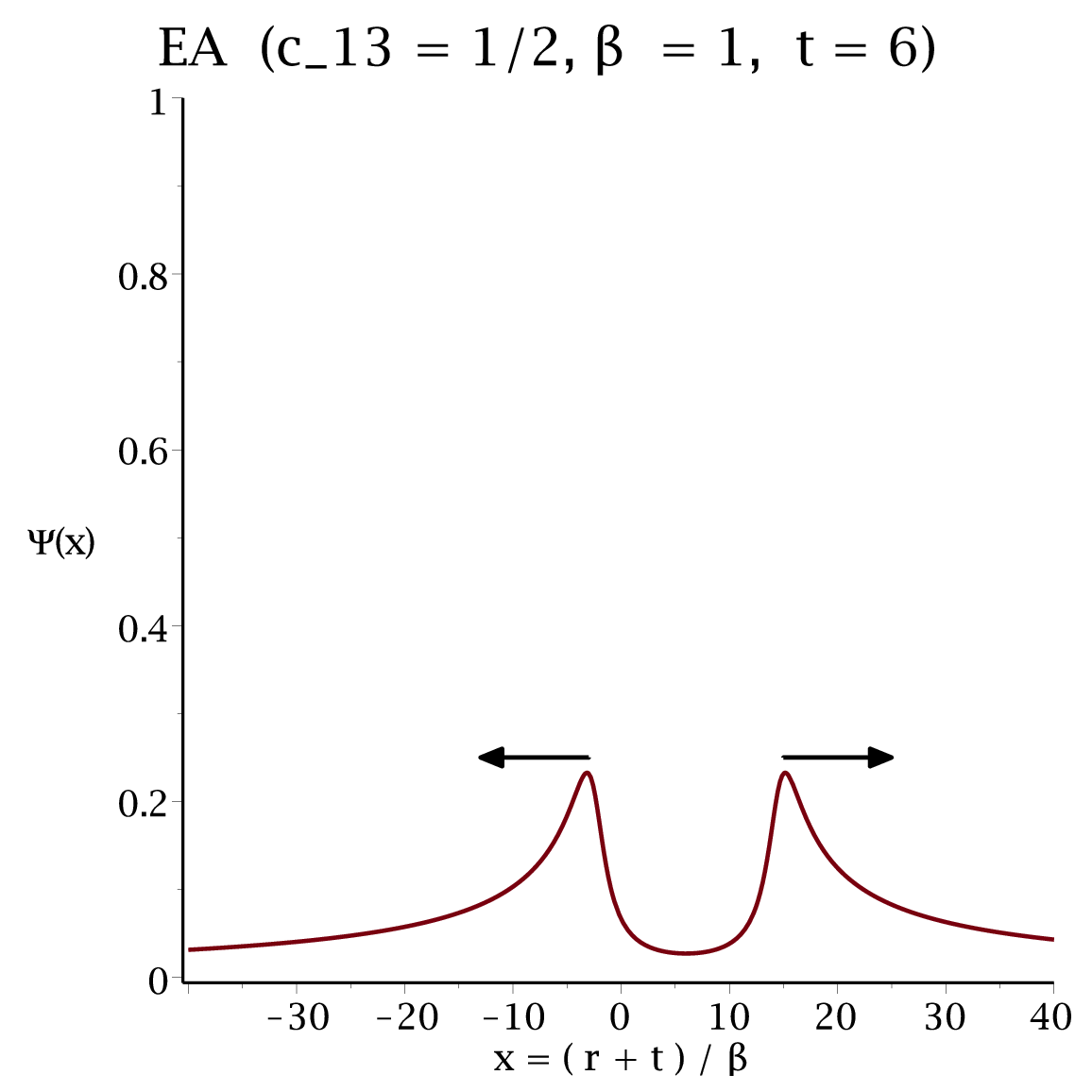}
	\includegraphics[width=7cm]{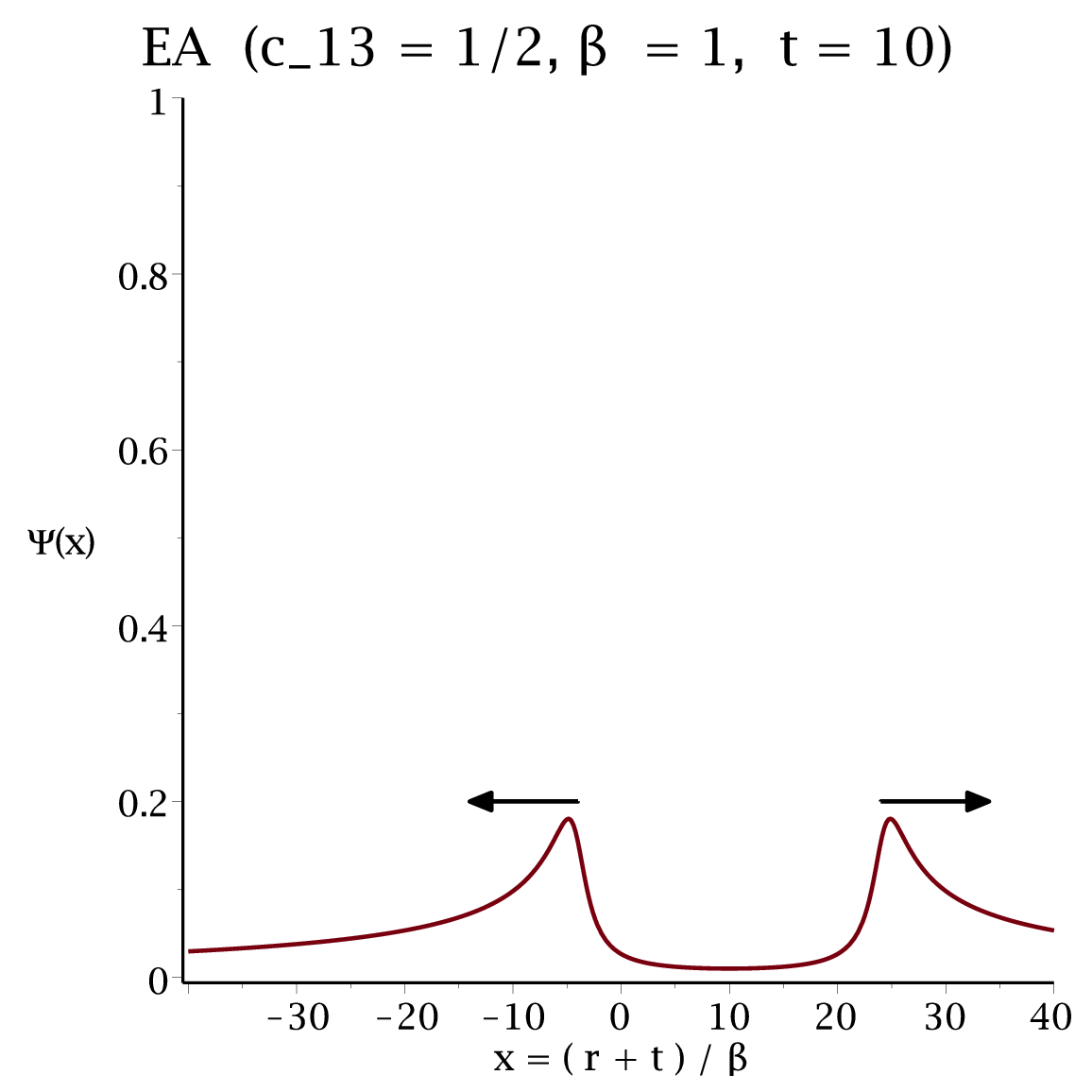}
	\caption{Wave component pulse $\Psi$ of width $\beta$, using $C_1=1$ in equation
		(\ref{Psirt}) and $\alpha=t$, $\beta=1$, $\gamma=\sqrt{1-c_{13}}\,r$ in equation (\ref{intpulse}), for four values of $t=0$, $t=2$, $t=6$ and $t=10$ and the parameters of Table \ref{table1}.}
	\label{fig2}
\end{figure}

\begin{figure}[!ht]
	\centering	
	\includegraphics[width=7cm]{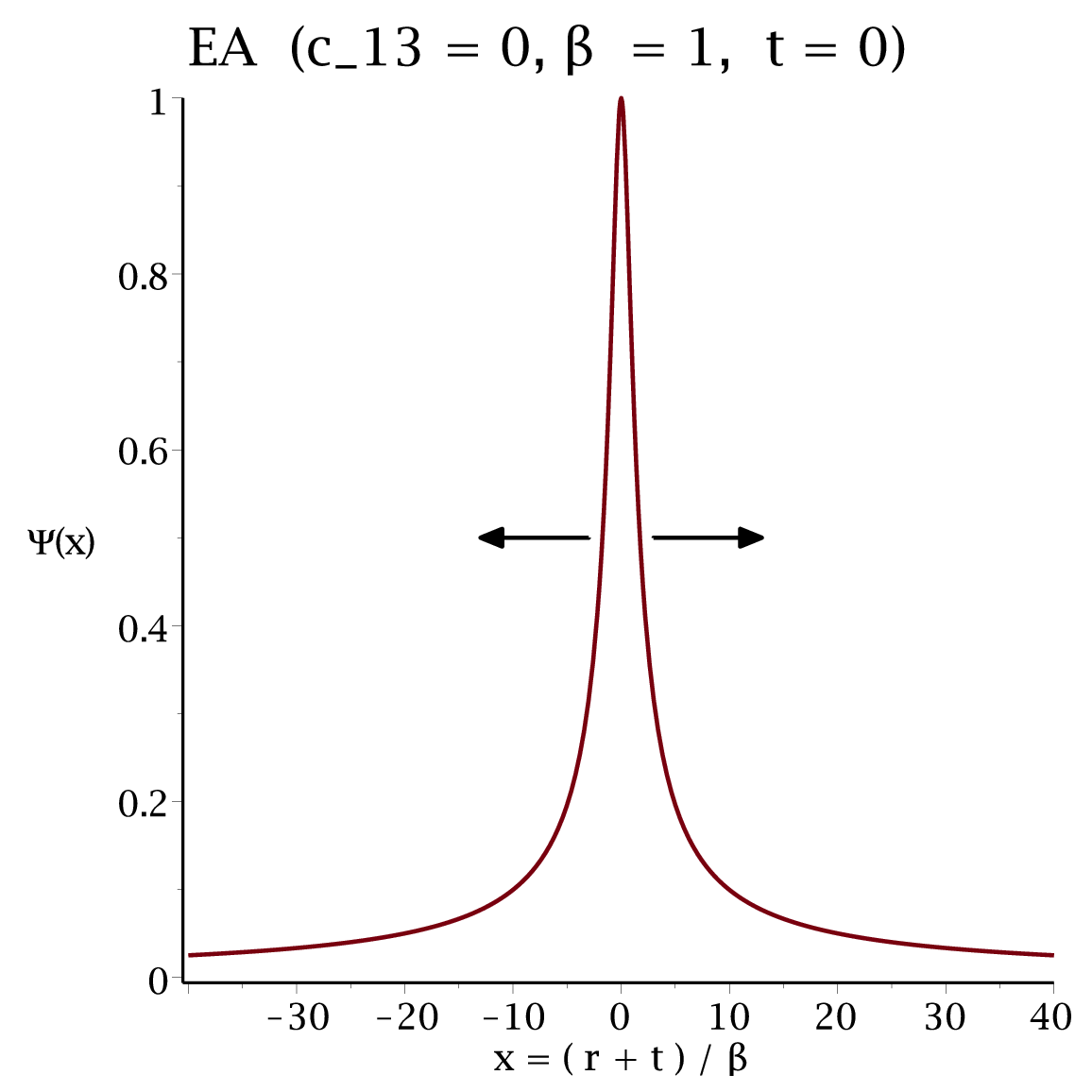}
	\includegraphics[width=7cm]{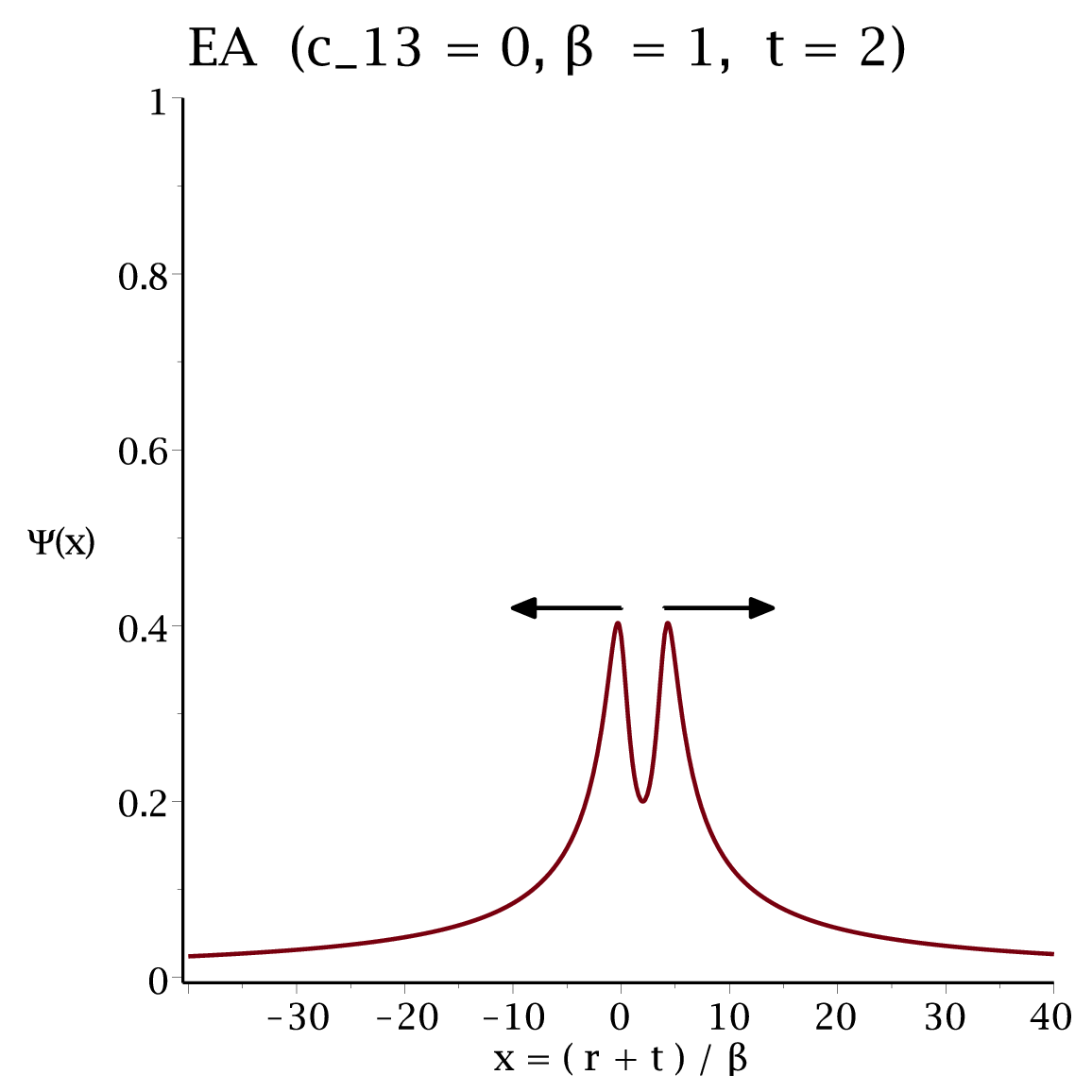}
	\includegraphics[width=7cm]{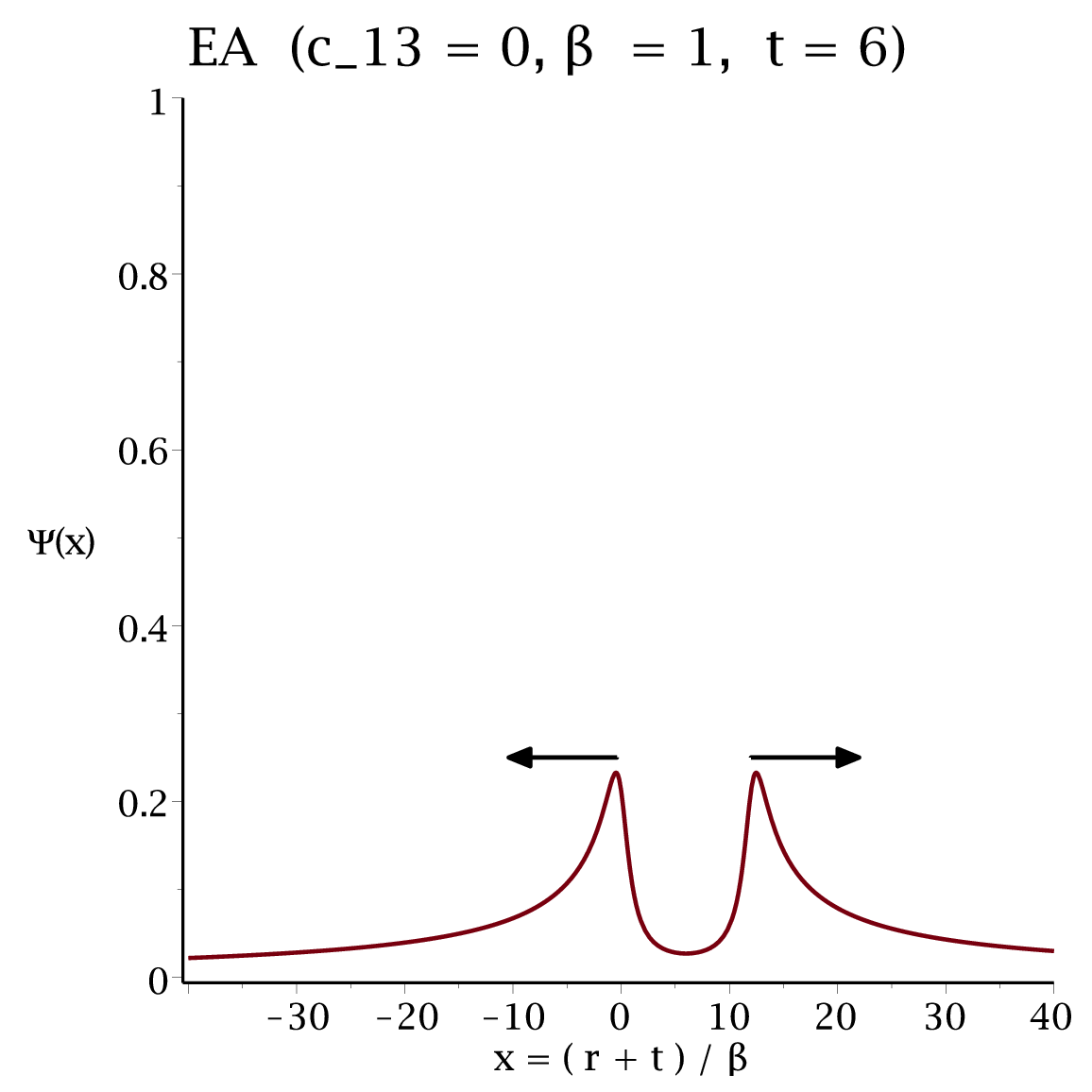}
	\includegraphics[width=7cm]{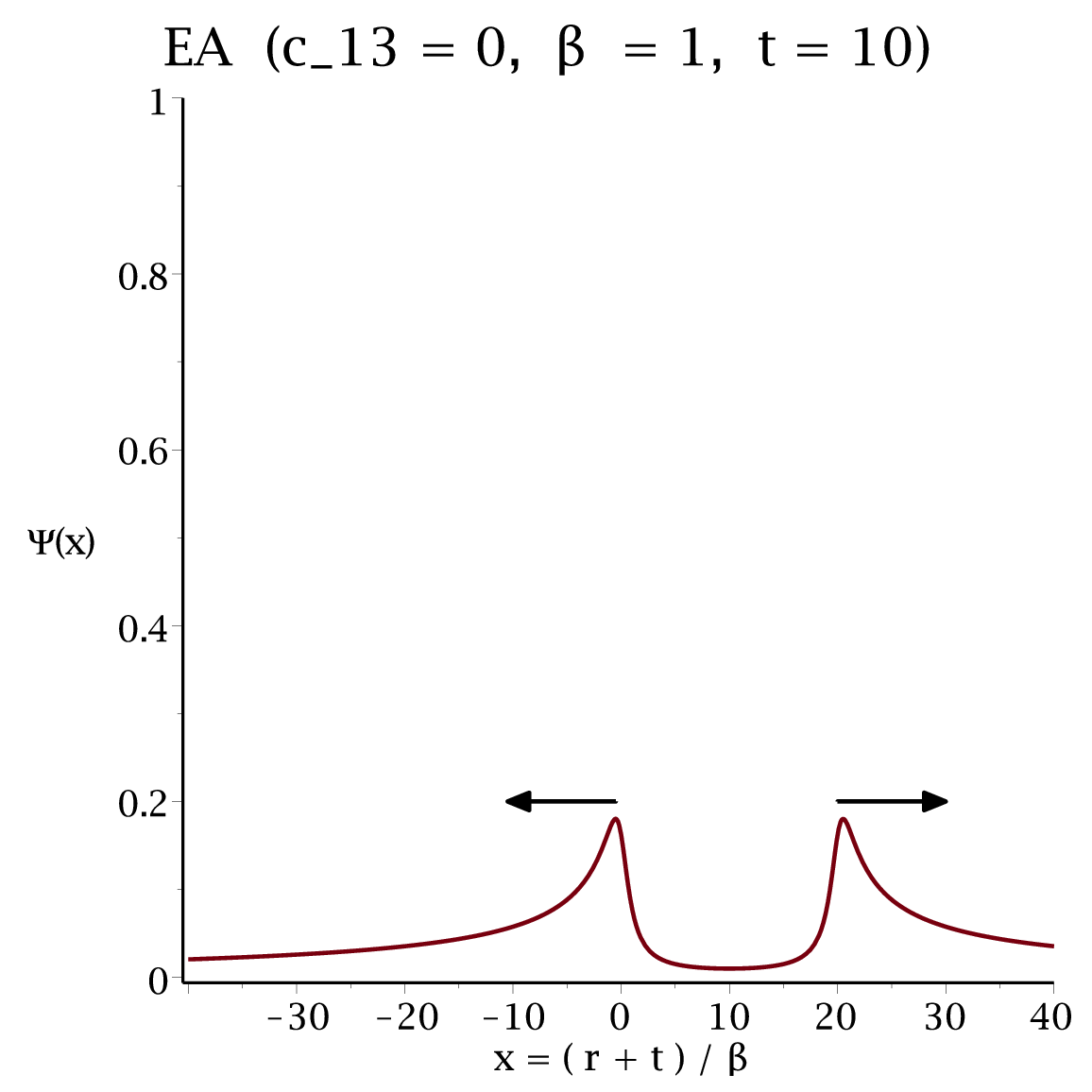}
	\caption{Wave component pulse $\Psi$ of width $\beta$, using $C_1=1$ in equation
		(\ref{Psirt}) and $\alpha=t$, $\beta=1$, $\gamma=\sqrt{1-c_{13}}\,r$ in equation (\ref{intpulse}), for four values of $t=0$, $t=2$, $t=6$ and $t=10$, and the parameters of Table \ref{table1}.}
	\label{fig3}
\end{figure}

\begin{figure}[!ht]
	\centering	
	\includegraphics[width=7cm]{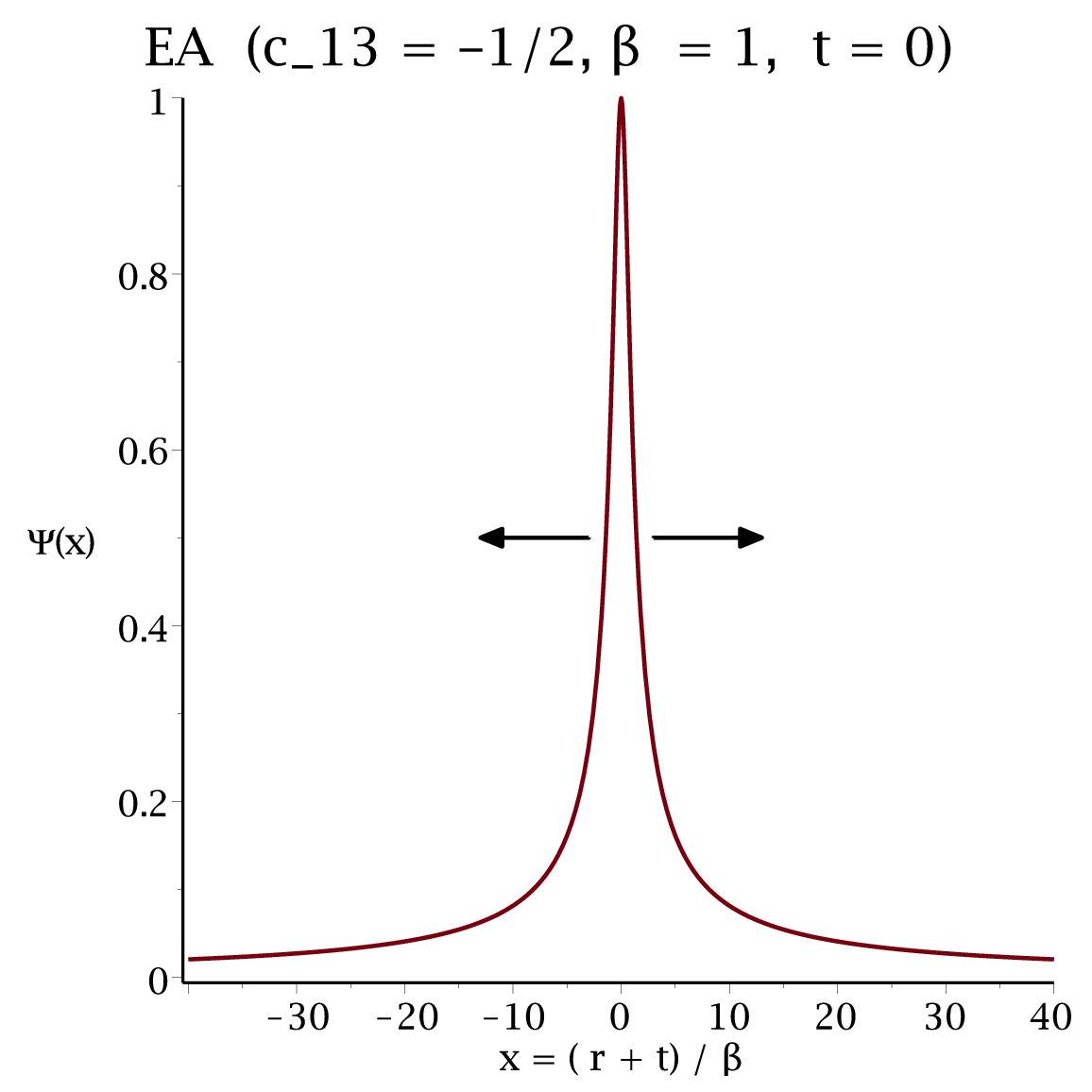}
	\includegraphics[width=7cm]{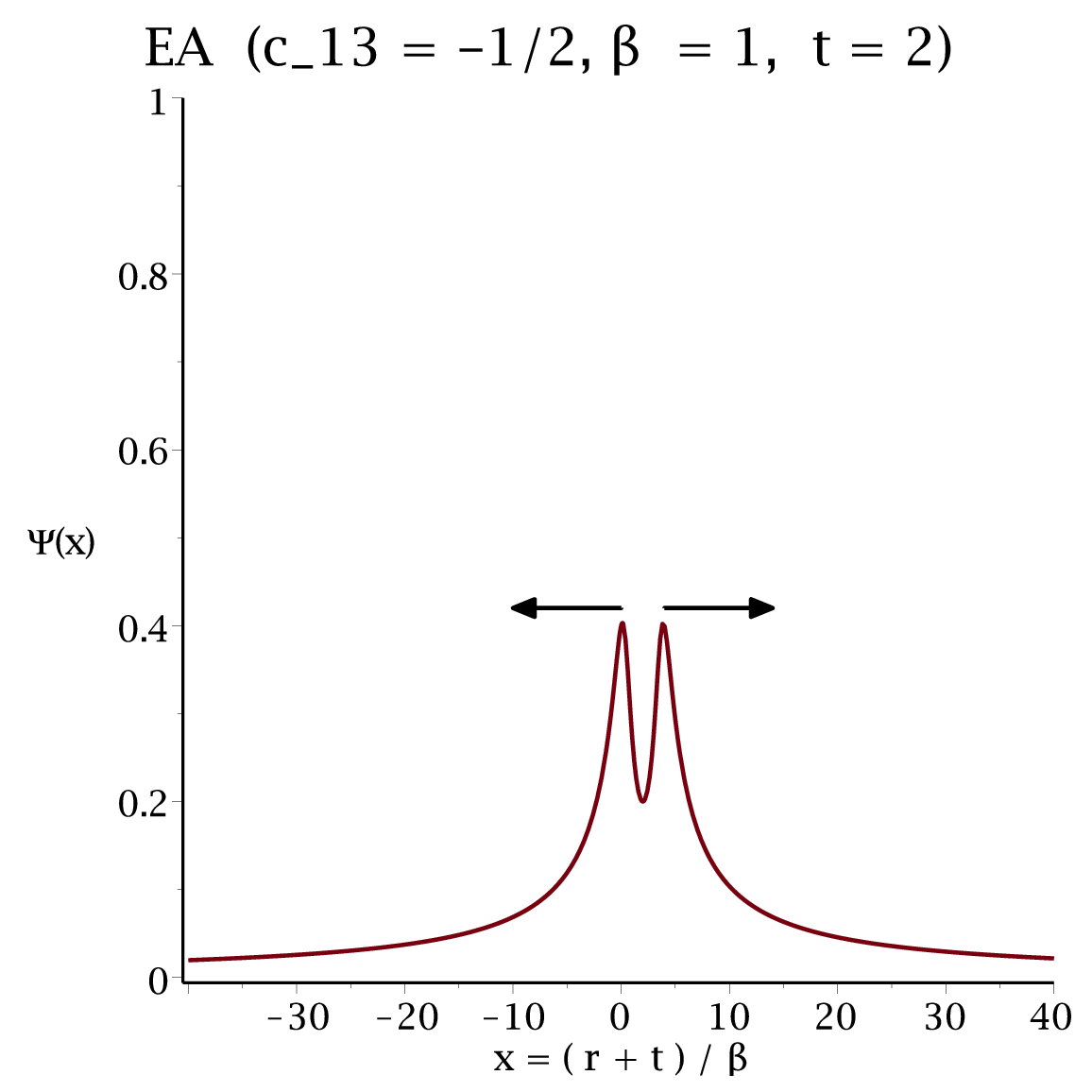}
	\includegraphics[width=7cm]{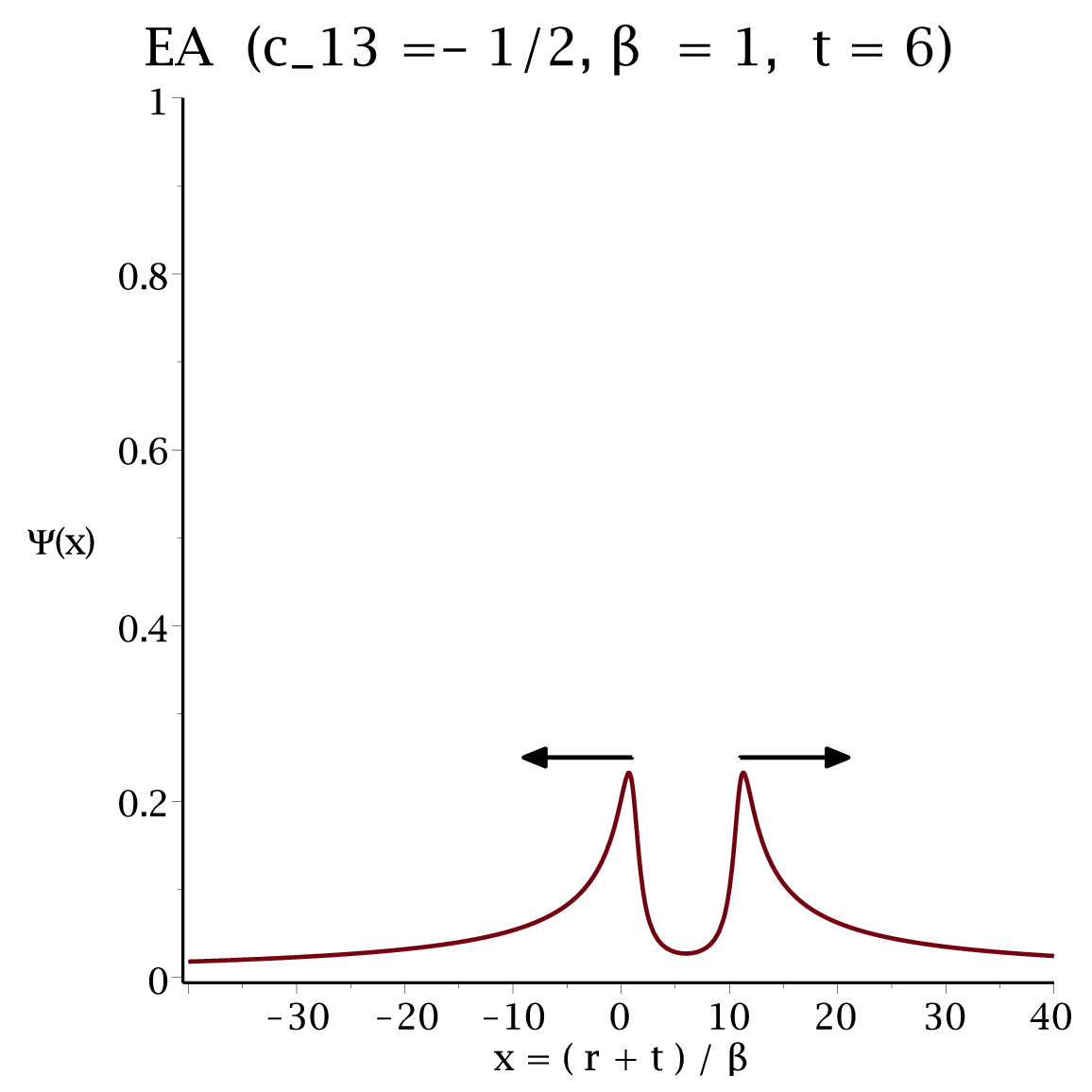}
	\includegraphics[width=7cm]{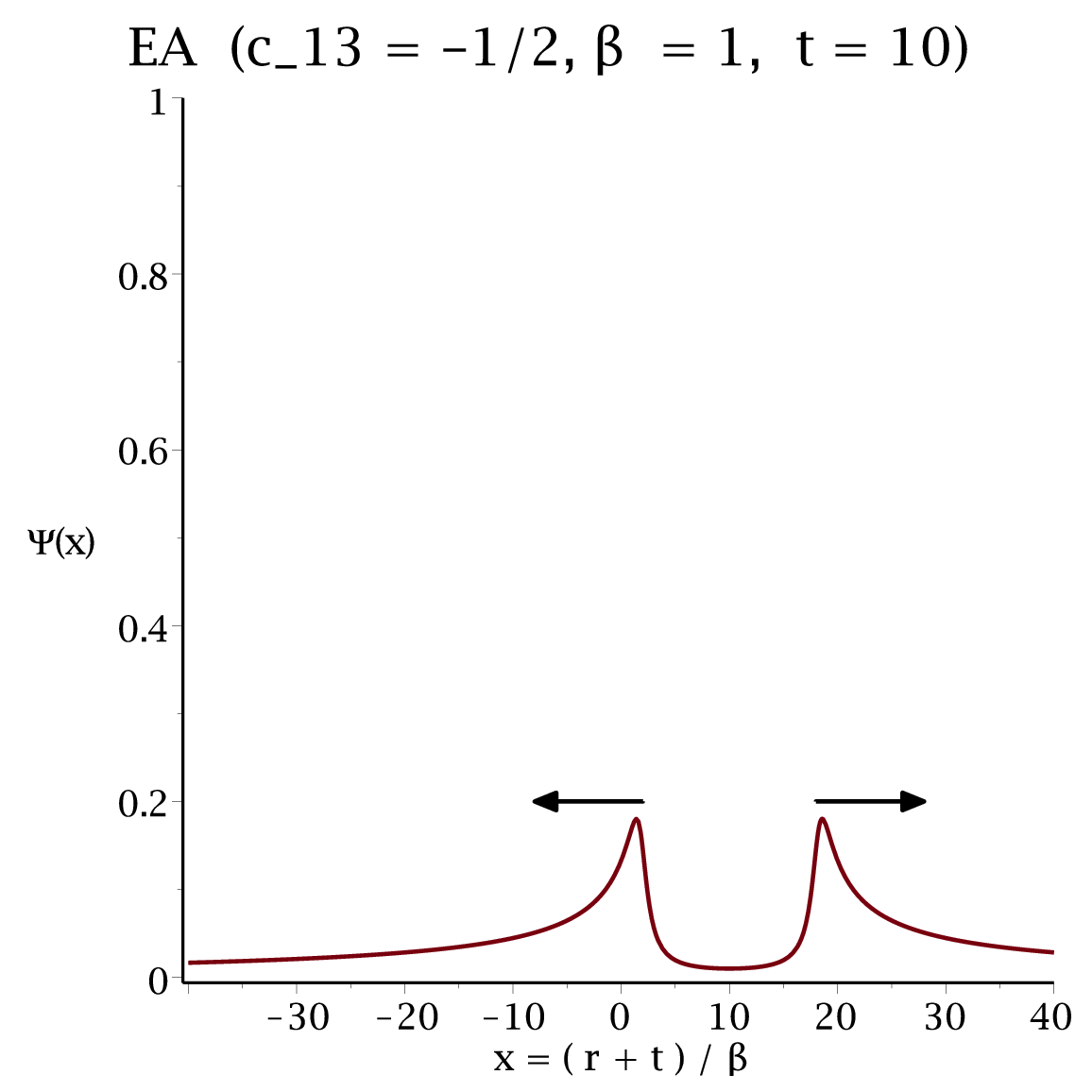}
	\caption{Wave component pulse $\Psi$ of width $\beta$, using $C_1=1$ in equation
		(\ref{Psirt}) and $\alpha=t$, $\beta=1$, $\gamma=\sqrt{1-c_{13}}\,r$ in equation (\ref{intpulse}), for four values of $t=0$, $t=2$, $t=6$ and $t=10$ and the parameters of Table \ref{table1}.}
	\label{fig4}
\end{figure}

\begin{figure}[!ht]
	\centering	
	\includegraphics[width=7cm]{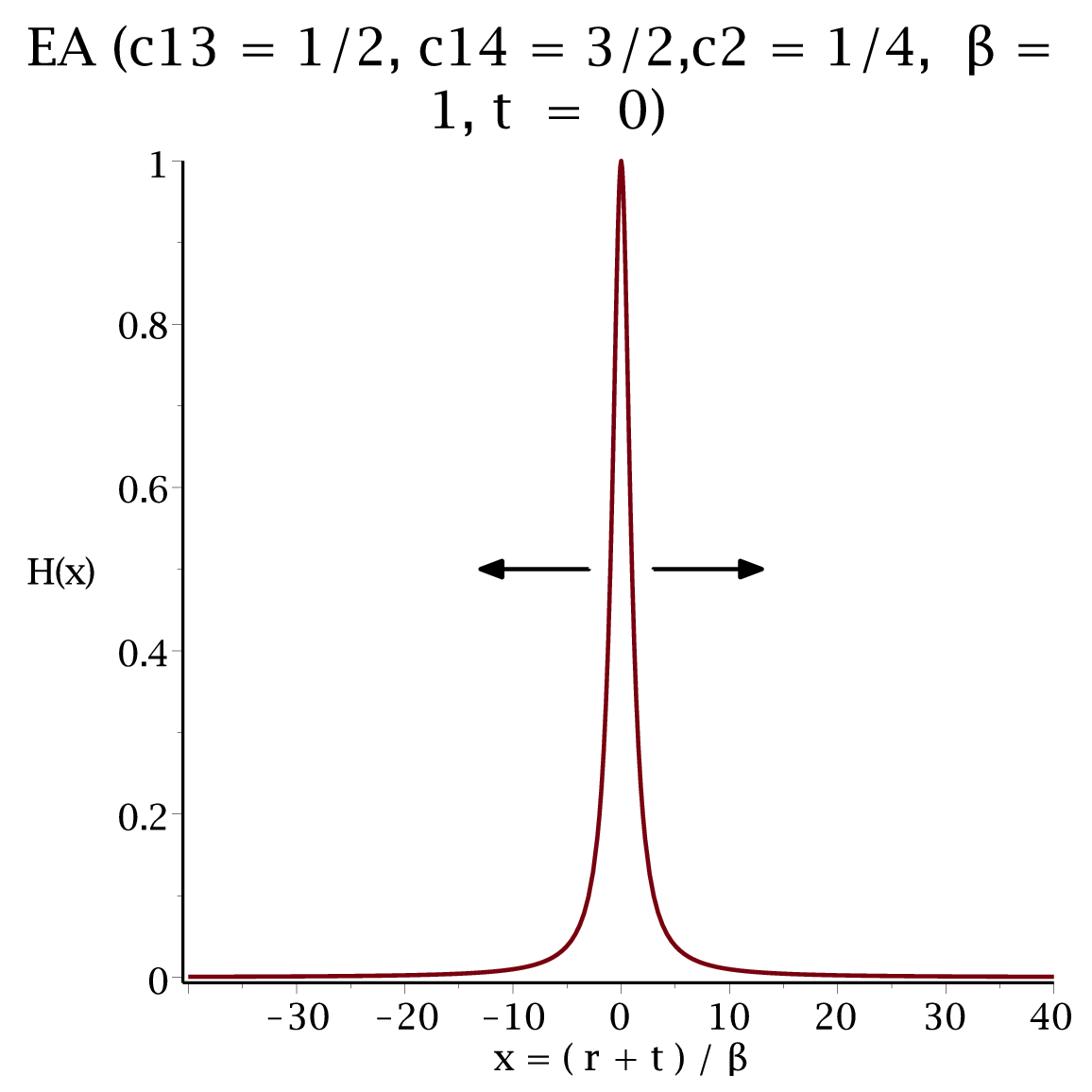}
	\includegraphics[width=7cm]{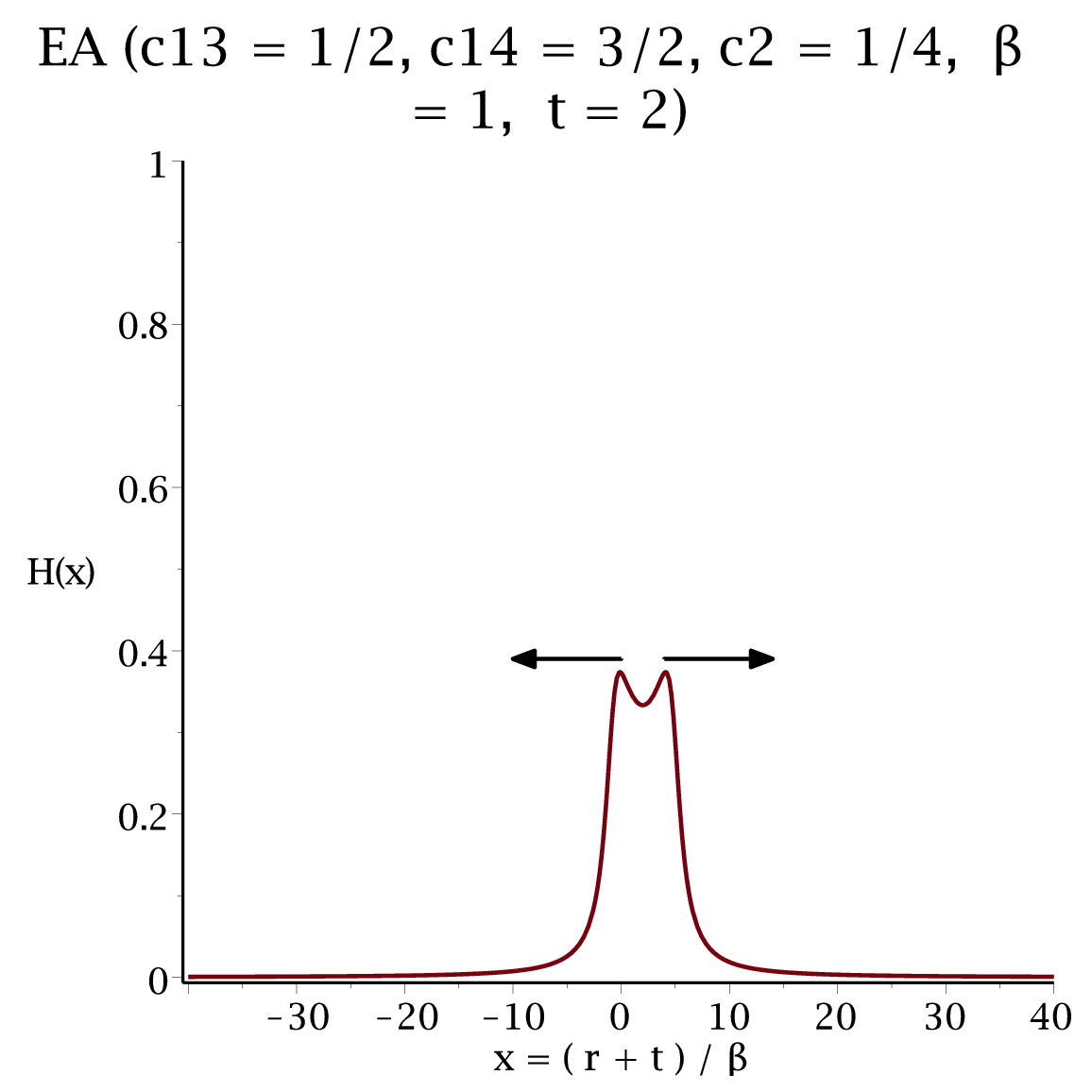}
	\includegraphics[width=7cm]{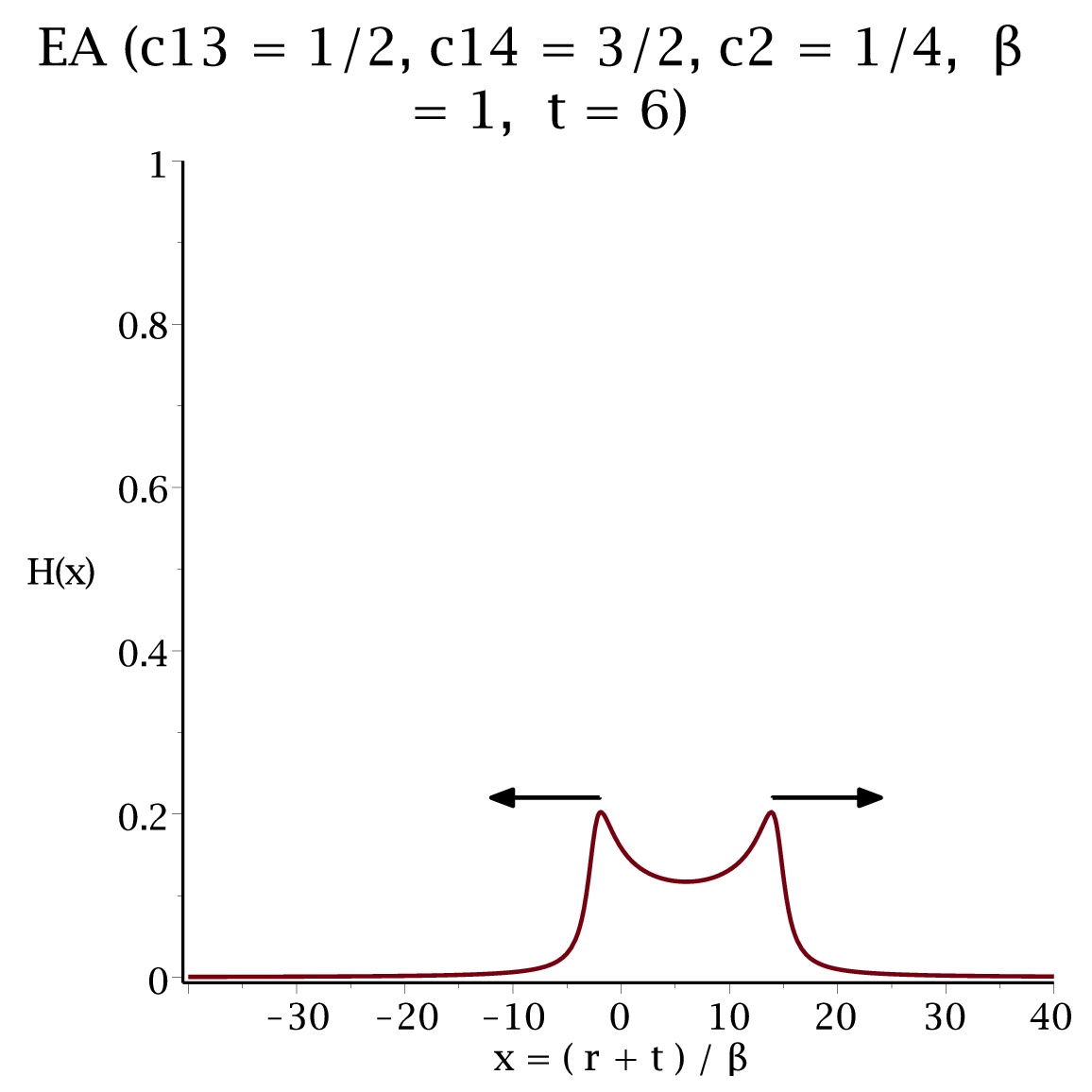}
	\includegraphics[width=7cm]{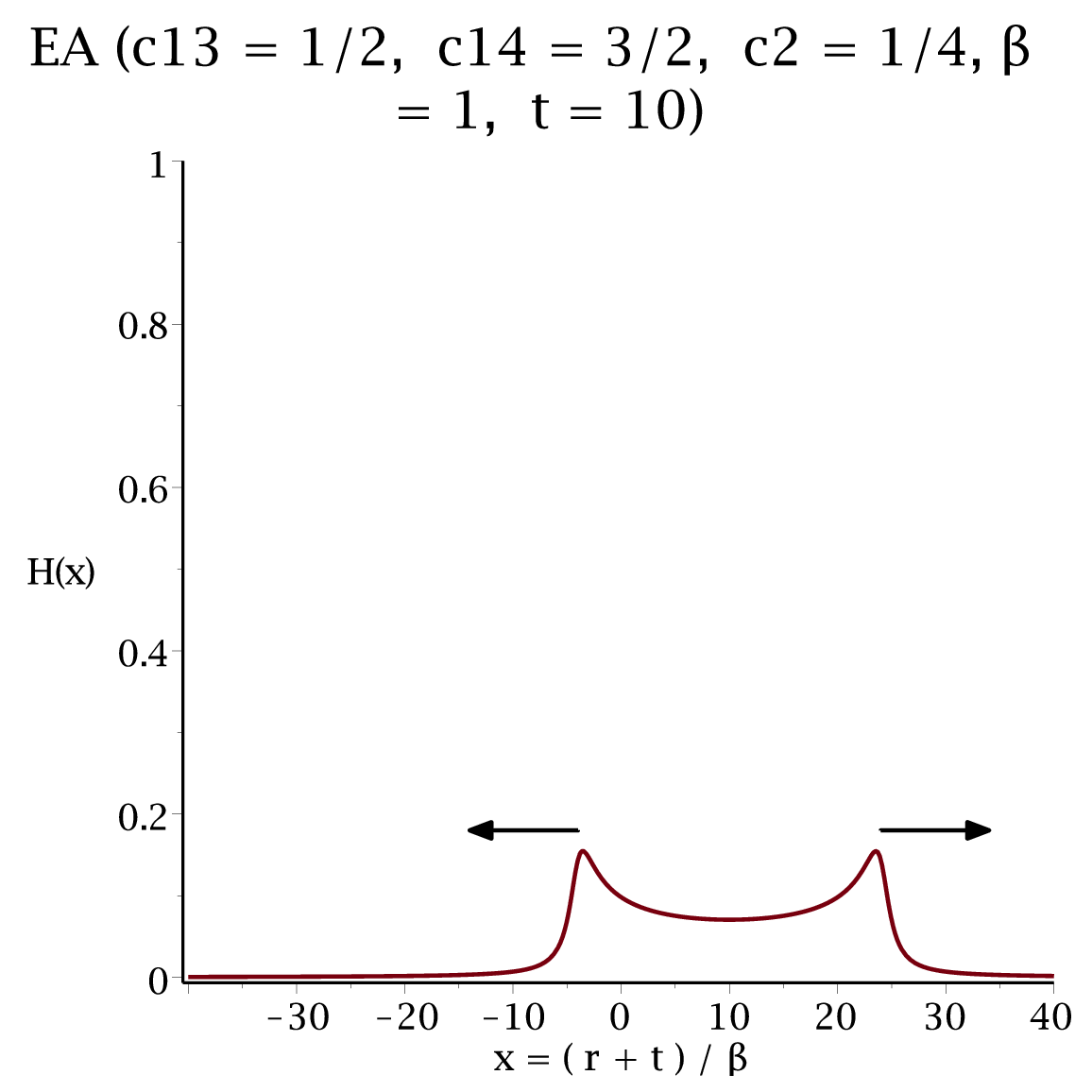}
	\caption{Wave component pulse $H$ of width $\beta$, using equation (\ref{Hrt}) and $\alpha={t\,\sqrt {c_{14}}}/{\sqrt {{c_{13}}+{c_2}}}$, $\beta=1$, $\gamma=r$ in equation (\ref{intpulse}), for four values of $t=0$, $t=2$, $t=6$ and $t=10$ and the parameters of Table \ref{table1}.}
	\label{fig5}
\end{figure}

\begin{figure}[!ht]
	\centering	
	\includegraphics[width=7cm]{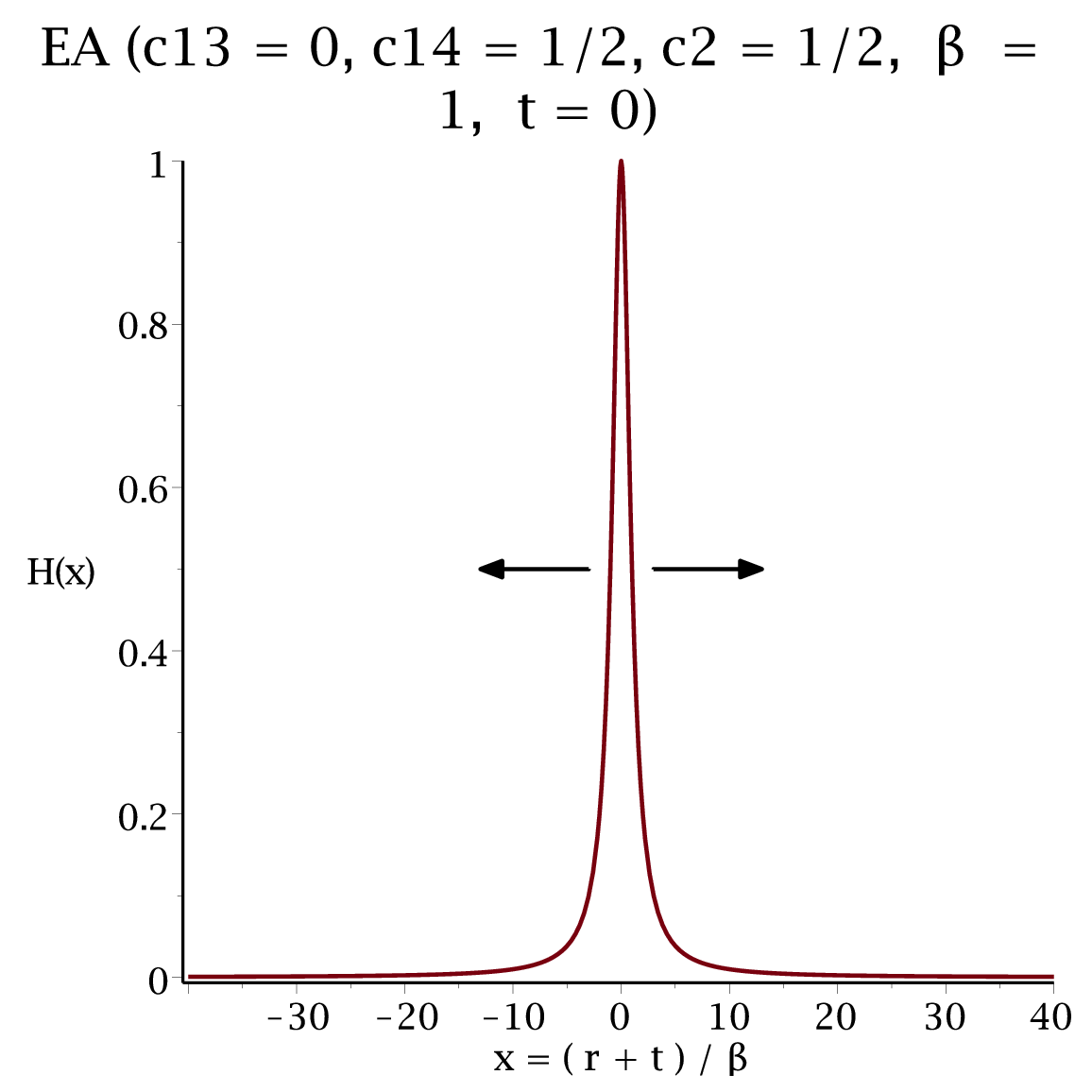}
	\includegraphics[width=7cm]{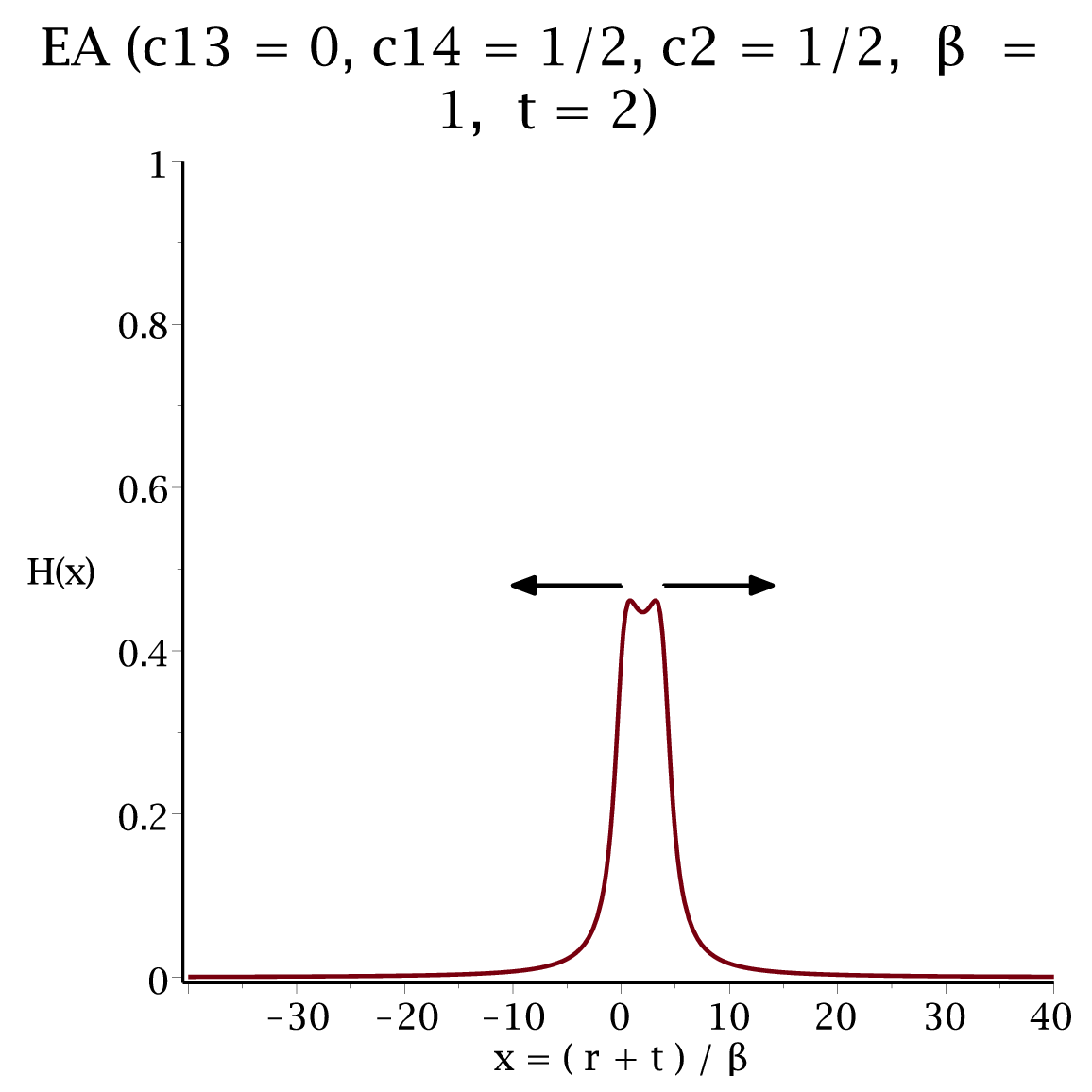}
	\includegraphics[width=7cm]{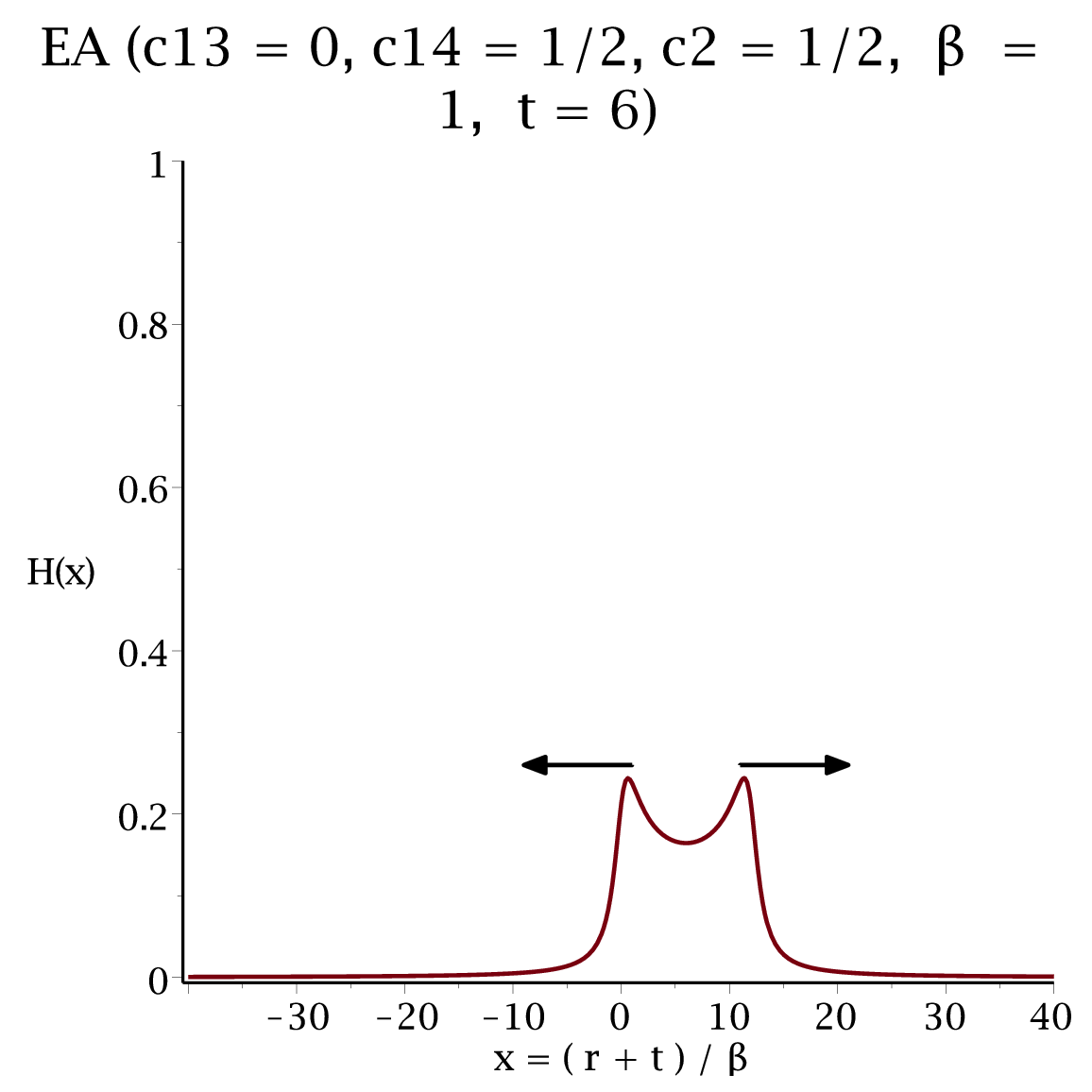}
	\includegraphics[width=7cm]{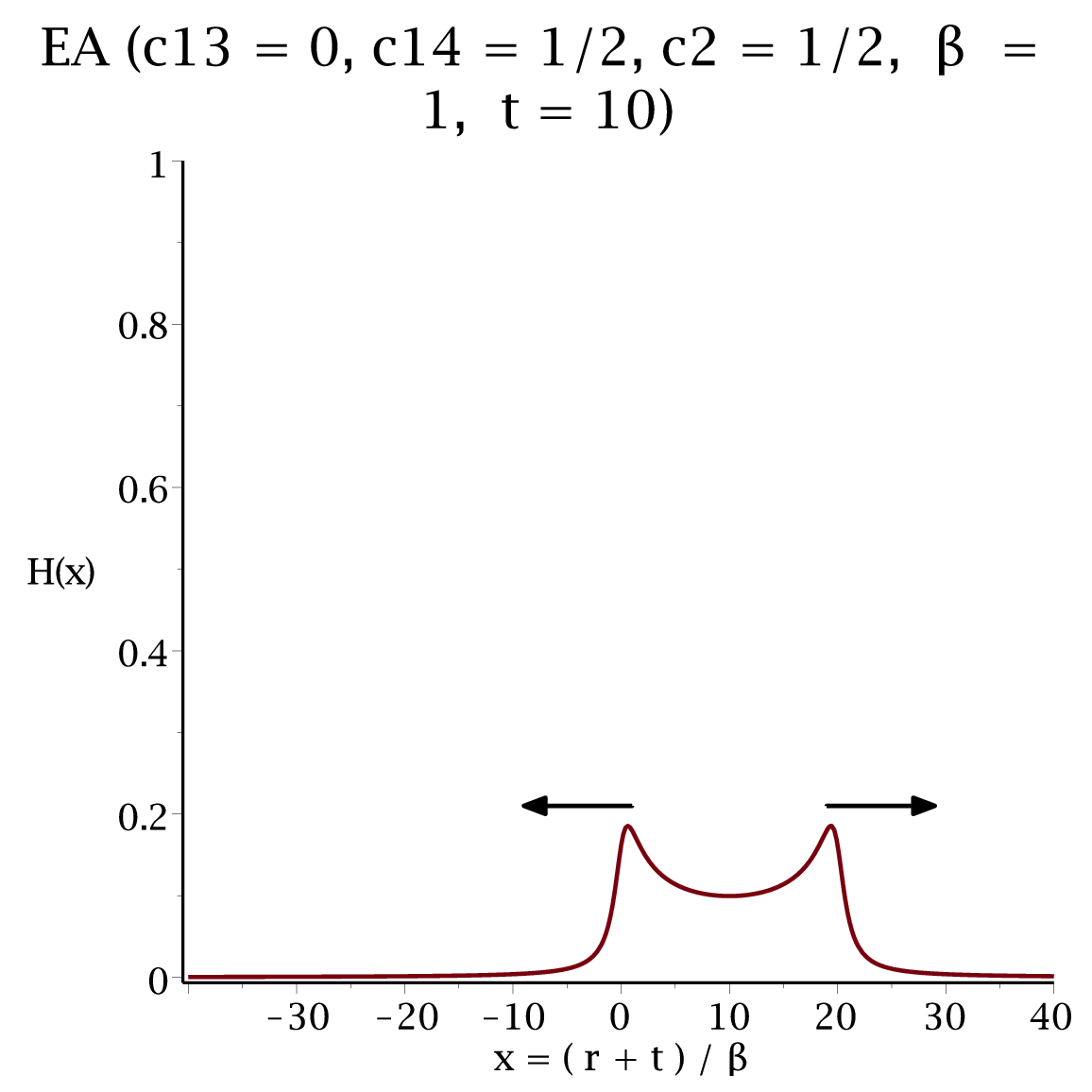}
	\caption{Wave component pulse $H$ of width $\beta$, using equation (\ref{Hrt}) and $\alpha={t\,\sqrt {c_{14}}}/{\sqrt {{c_{13}}+{c_2}}}$, $\beta=1$, $\gamma=r$ in equation (\ref{intpulse}), for four values of $t=0$, $t=2$, $t=6$ and $t=10$, and the parameters of Table \ref{table1}.}
	\label{fig6}
\end{figure}

\begin{figure}[!ht]
	\centering	
	\includegraphics[width=7cm]{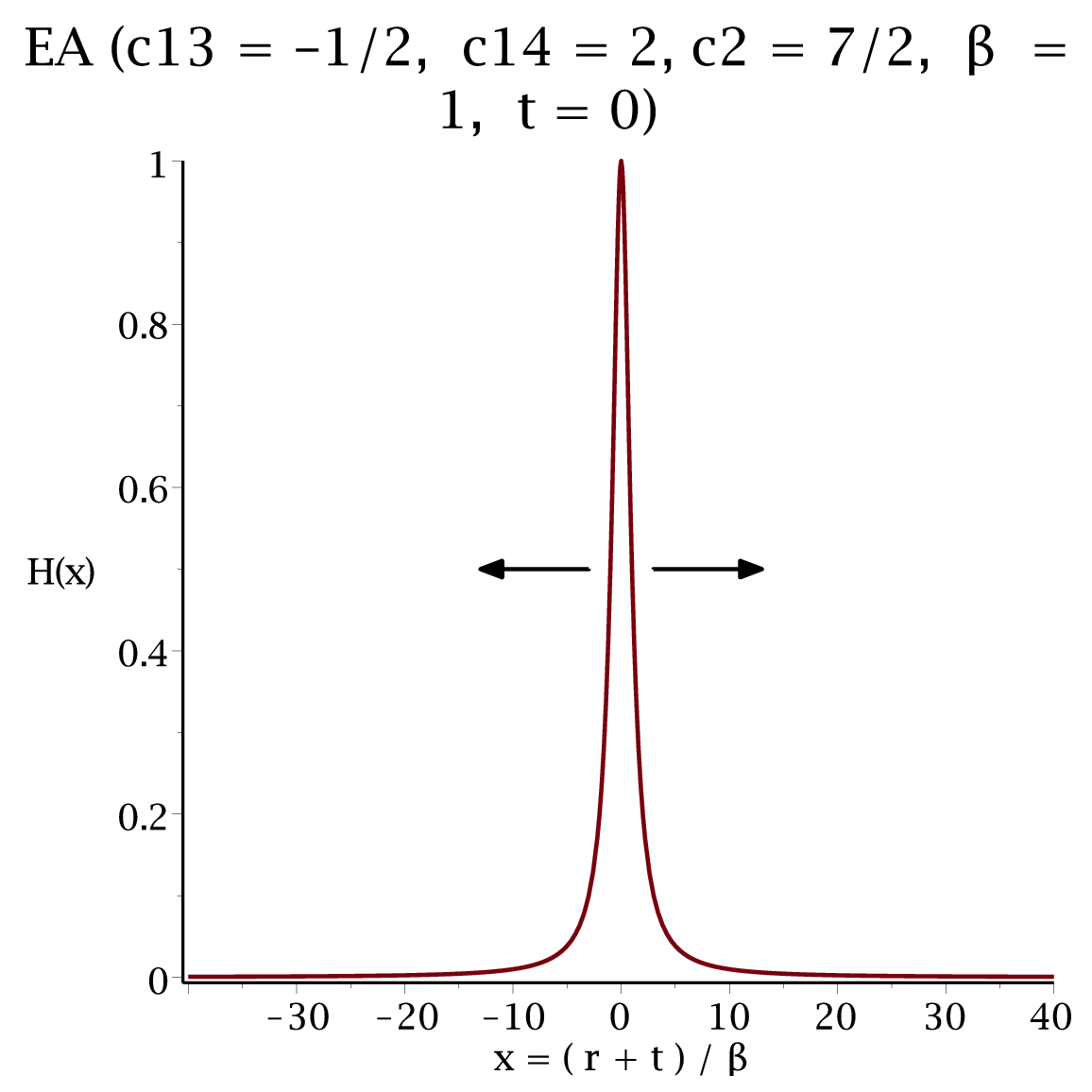}
	\includegraphics[width=7cm]{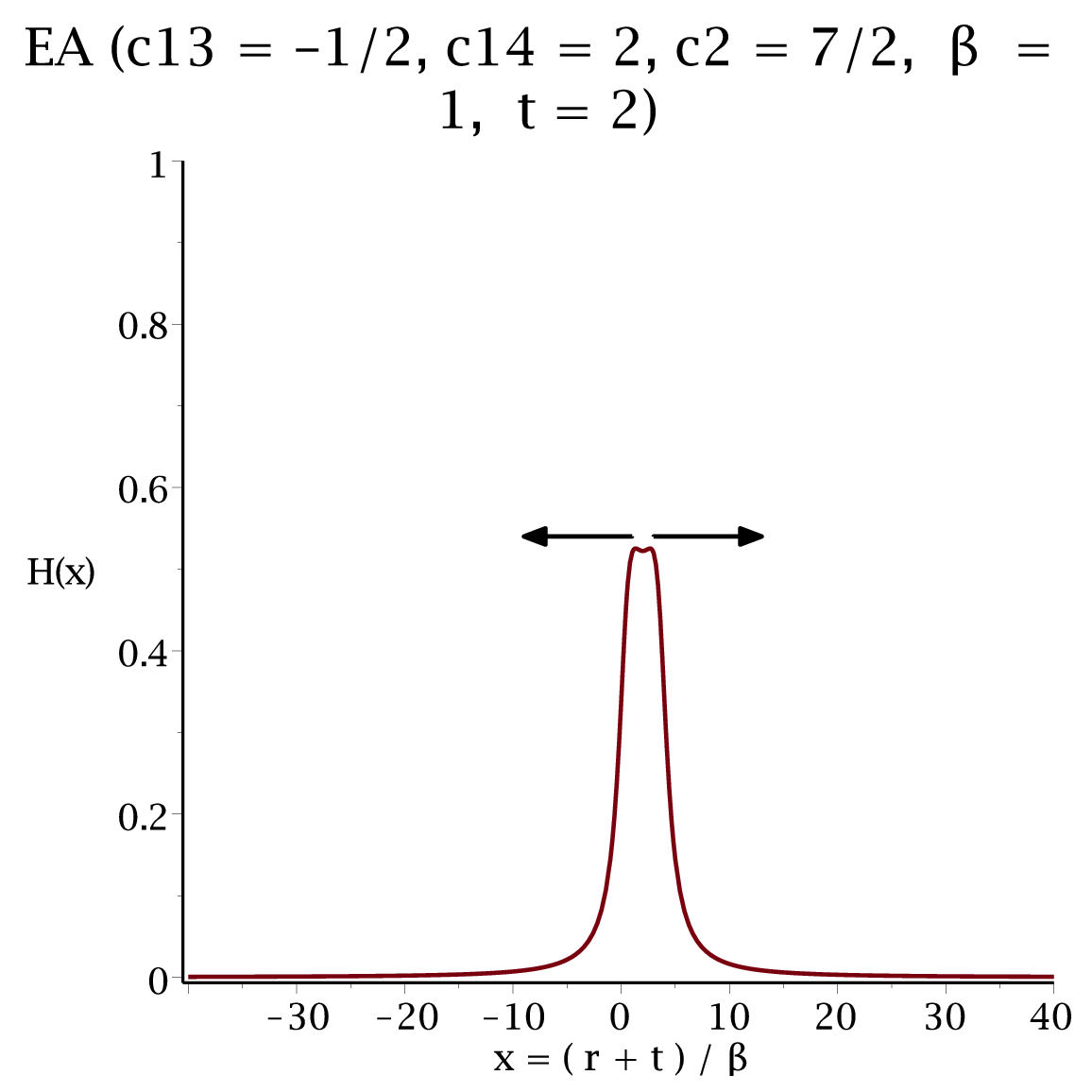}
	\includegraphics[width=7cm]{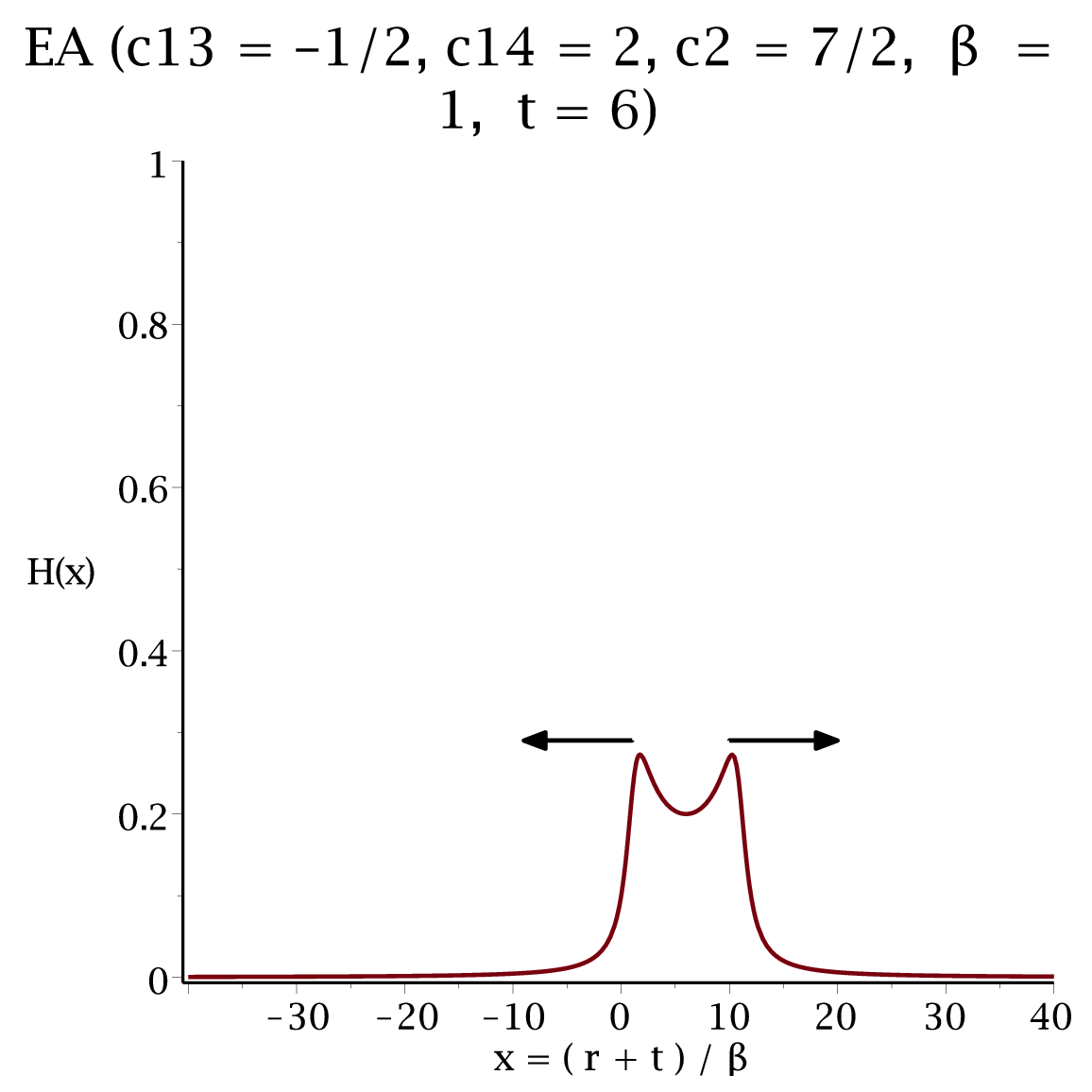}
	\includegraphics[width=7cm]{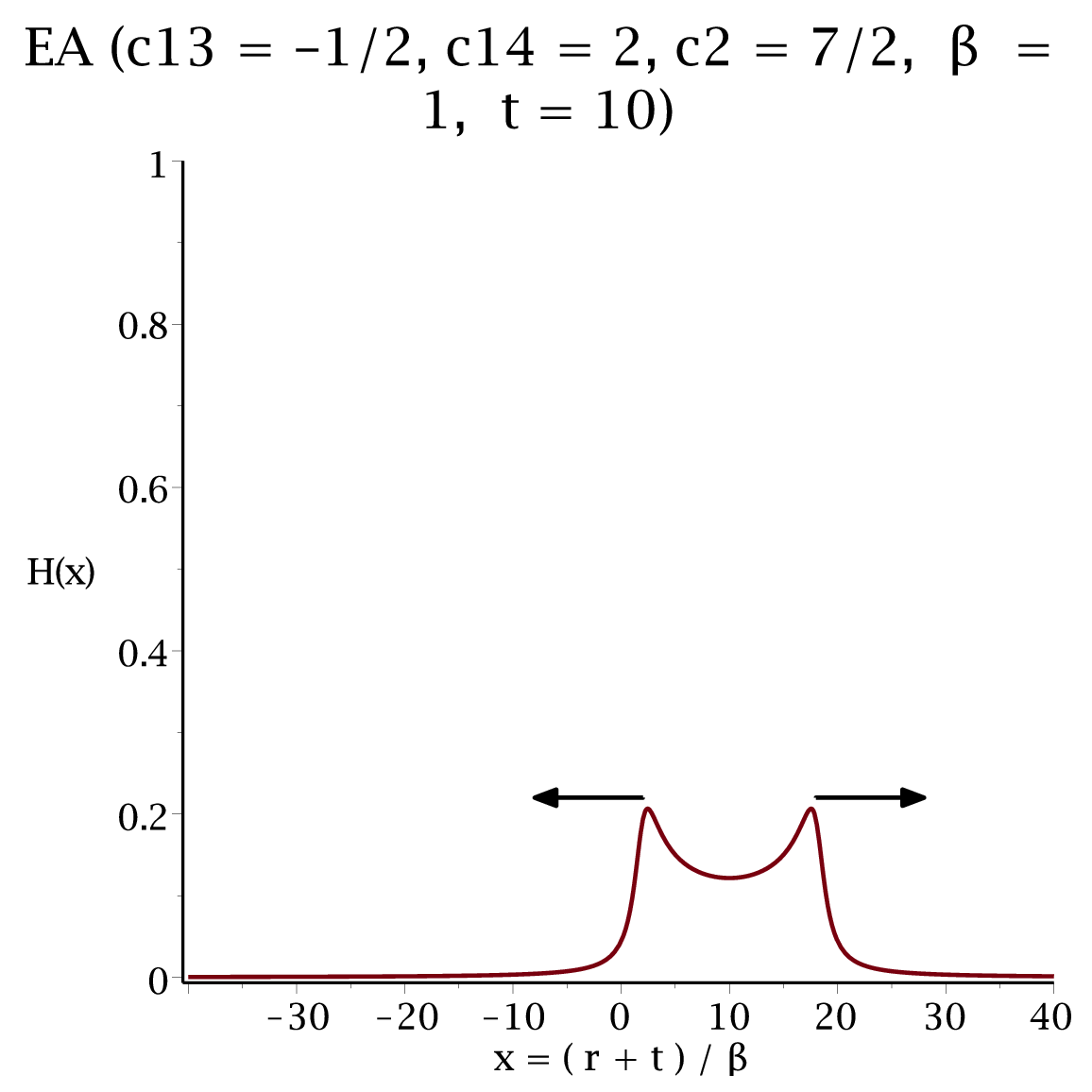}
	\caption{Wave component pulse $H$ of width $\beta$, using	 in equation (\ref{Hrt}) and $\alpha={t\,\sqrt {c_{14}}}/{\sqrt {{c_{13}}+{c_2}}}$, $\beta=1$, $\gamma=r$ in equation (\ref{intpulse}), for four values of $t=0$, $t=2$, $t=6$ and $t=10$ and the parameters of Table \ref{table1}.}
	\label{fig7}
\end{figure}

\section{Comparison of $\bf \Psi(r,t)$ and $\bf H(r,t)$ in EA and GR}

Assuming (\ref{ds2}) and $c_1=c_2=c_3=c_4=0$ we compute the different terms in the field equations (\ref{fieldeqs}), giving 
\bqn
G_{00} &=& 2 H_{1}-2 \Psi_{1}^2 r-2 \Psi_{0}^2 r =0,
\lb{G00a}
\eqn
\bqn
G_{01}&=&H_{0}+2 \Psi_{1} \Psi_{0} r=0,
\lb{G01a}
\eqn
\bqn
G_{11} &=&2 H_{1}-2 \Psi_{1}^2 r-2 \Psi_{0}^2 r =0,
\lb{G11a}
\eqn
\bqn
G_{22} &=&2 \Psi_{1}^2-2 H_{00}+2 H_{11}-2 \Psi_{0}^2 =0,
\lb{G22a}
\eqn
\bqn
G_{33} &=&4 \Psi_{1}-2 \Psi_{1}^2 r+2 \Psi_{0}^2 r-2 r H_{11}+4 \Psi_{11} r-4 r \Psi_{00}+2 r H_{00} =0.
\lb{G33a}
\eqn

The wave equation for GR is obtained using (\ref{G33a}) and (\ref{G22a}), thus
\bqn
{\Psi_{00}}= \left( \Psi_{11}+ \frac{\Psi_{1}}{r} \right).
\lb{wavea}
\eqn

Substituting $\Psi_1$ from equation (\ref{wavea}) into equations (\ref{G01a}) and 
(\ref{G00a}) we get
\bq
{H_0}-2\,{\Psi_0}\,{r}^{2}{\Psi_{00}}+2\,{\Psi_0}\,{r}^{2}{
	\Psi_{11}}=0,
\lb{H0a}
\eq
\bq
{H_1}-{r}^{3}{{\Psi_{00}}}^{2}+2\,{r}^{3}{\Psi_{00}}\,{\Psi_{11}}-{
	{\Psi_{11}}}^{2}{r}^{3}-{{\Psi_0}}^{2}r=0,
\lb{H1a}
\eq
This must be solved to obtain $H(r,t)$.
Notice that these two equations are not equivalent to the equation (\ref{bound}) in EA.
In GR these equations come from the components $G_{01}$ and $G_{00}$, while in EA
the equation (\ref{bound}) comes from equations (\ref{G11}) and (\ref{G00}).

Solving equation (\ref{wavea}) we get
\bq
\Psi(r,t)= \left[ {C_{13}}\,{J_0\left({\mathrm w}r\right)}+{C_{14}}\,
{Y_0\left({\mathrm w}r\right)} \right] \cos \left( {\mathrm w}t \right) .
\lb{Psirta}
\eq
We will assume $C_{14}=0$ since $Y_0$ is singular at $r=0$ and $C_{13}=1$
for simplicity. Thus,
\bq
\Psi(r,t)= {J_0\left({\mathrm w}r\right)} \cos \left( {\mathrm w}t \right).
\lb{Psirtb}
\eq

\begin{figure}[!ht]
	\centering	
	\includegraphics[width=7cm]{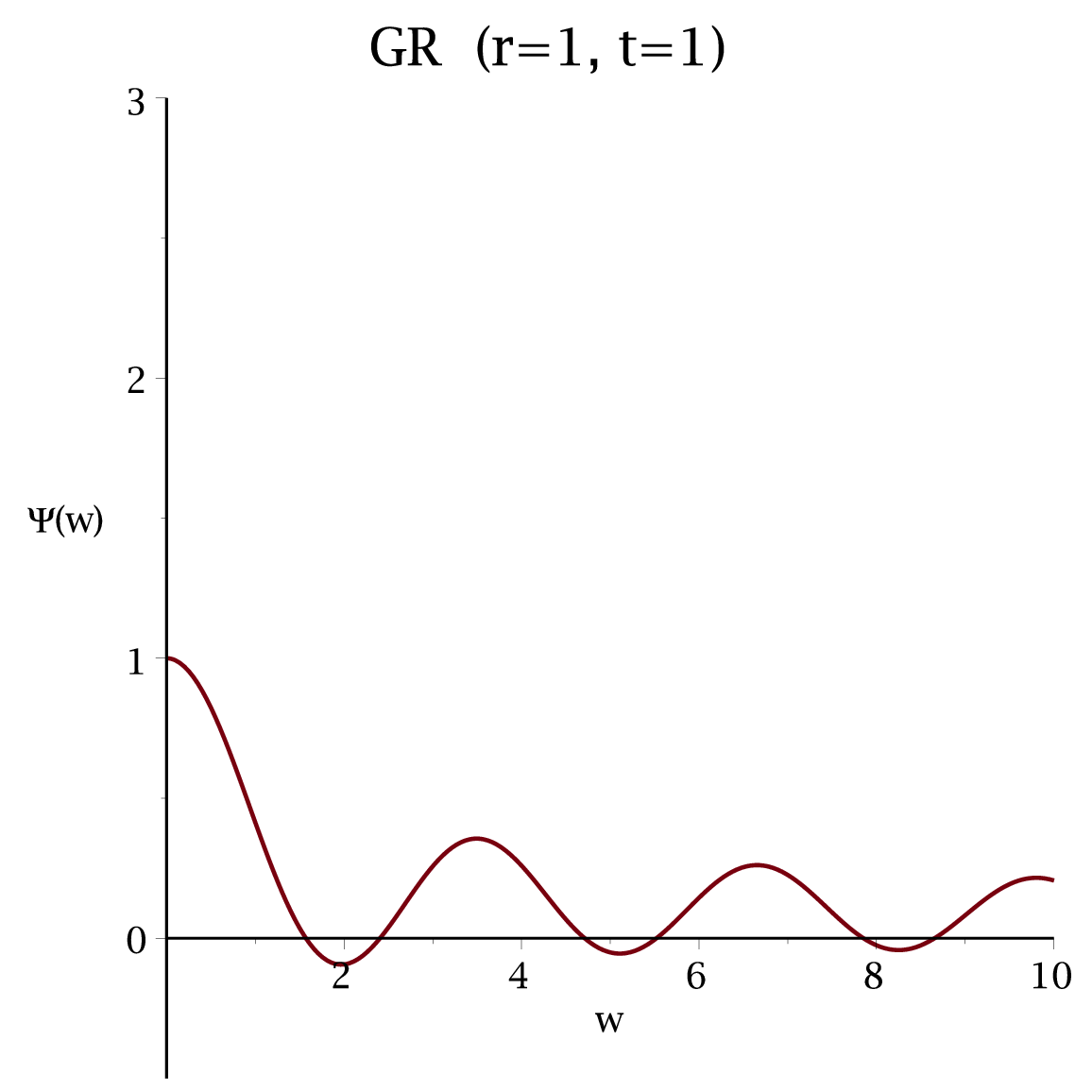}
	\includegraphics[width=7cm]{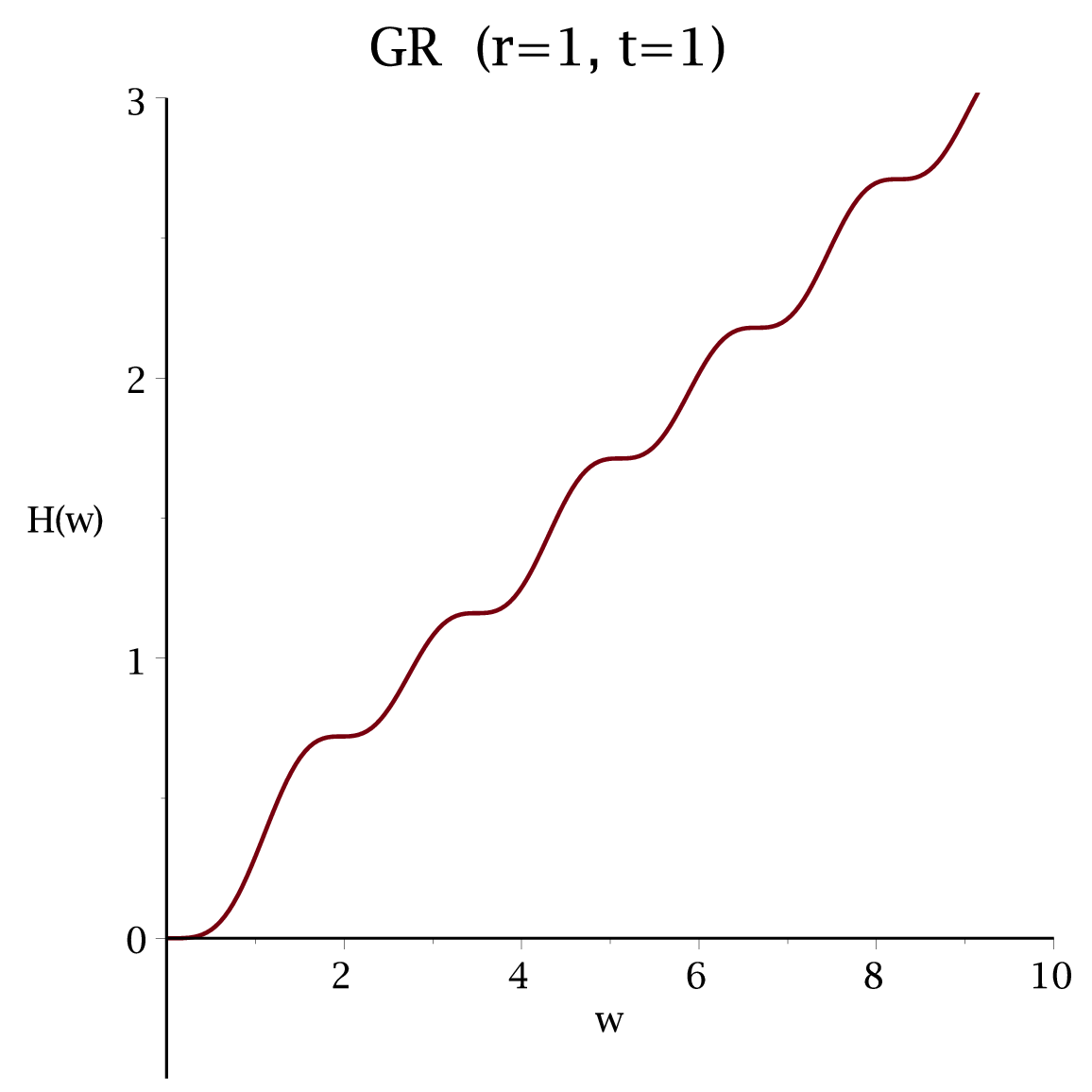}
	\caption{Wave components  $\Psi({\mathrm w})$ and $H(\omega)$ using $r=1$ and $t=1$ 
		in equations (\ref{Psirtb}) and (\ref{Hrtb}).} 
	\label{fig8}
\end{figure}

Substituting $\Psi$ from (\ref{Psirta}) into (\ref{H0a}) and (\ref{H1a}) and solving
both equations, we obtain
\bqn
H(r,t)&=&
-\frac{1}{2}\,{\mathrm w}r \left\{ 2\, \cos \left( {\mathrm w}t \right) ^{2}
J_0\left({\mathrm w}r\right)J_1\left({\mathrm w}r\right)-\right.\nb\\
&&\left. {\mathrm w}r \left[ J_0\left({\mathrm w}r\right) ^{2}+  
J_1\left({\mathrm w}r\right) ^{2} \right] \right\}.
\lb{Hrtb}
\eqn
See Figure \ref{fig8} for $\Psi({\mathrm w})$ and $H(\omega)$. 

Notice that the last term of this equation is not periodic in time, turning the solution of the gravitational wave into a quasi-periodic one. This differs entirely from the EA gravitational wave, where the metric functions $\Psi(r,t)$ and $H(r,t)$ are both completely periodic in time.

\begin{figure}[!ht]
	\centering	
	\includegraphics[width=7cm]{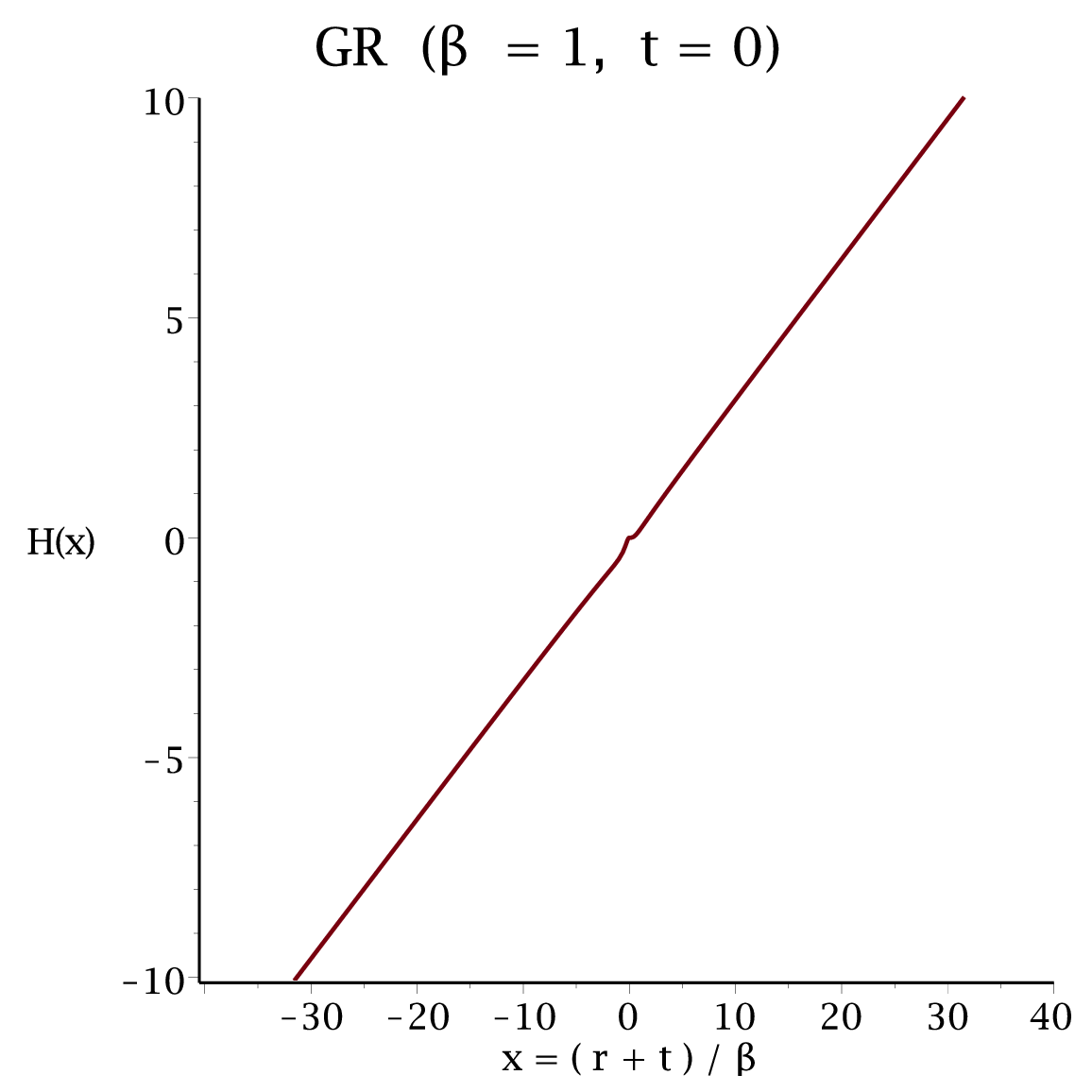}
	\caption{Wave component pulse $H$ of width $\beta$, using  equation (\ref{Hrt}) and $\alpha=t$, $\beta=1$, $\gamma=r$ in equation (\ref{intpulse}), for four values of $t=0$, $t=2$, $t=6$ and $t=10$. Fitting a straight line, we get $H_s \approx 0.1862+ 0.3186\,x$.} 
	\label{fig9}
\end{figure}

\begin{figure}[!ht]
	\centering	
	\includegraphics[width=7cm]{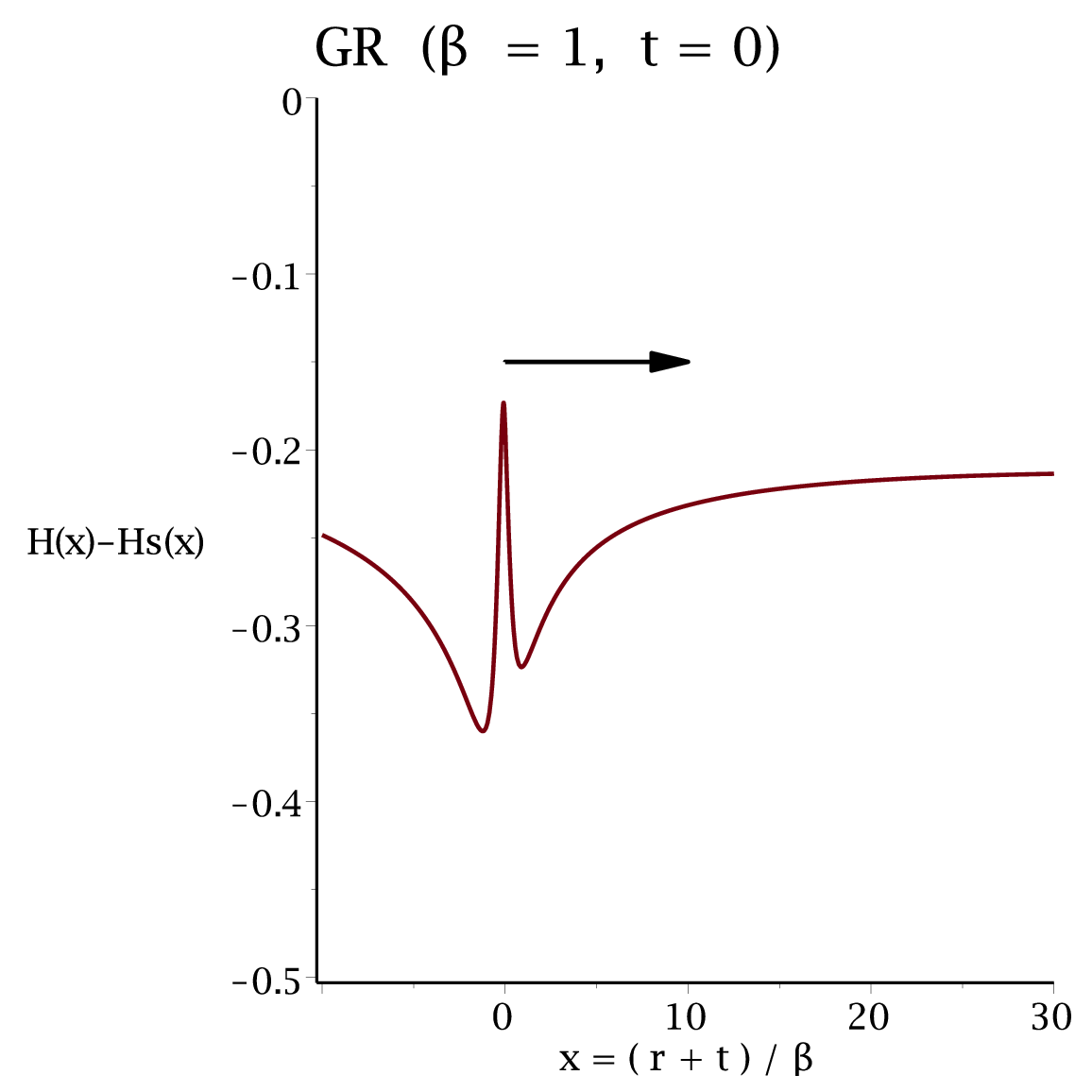}
	\includegraphics[width=7cm]{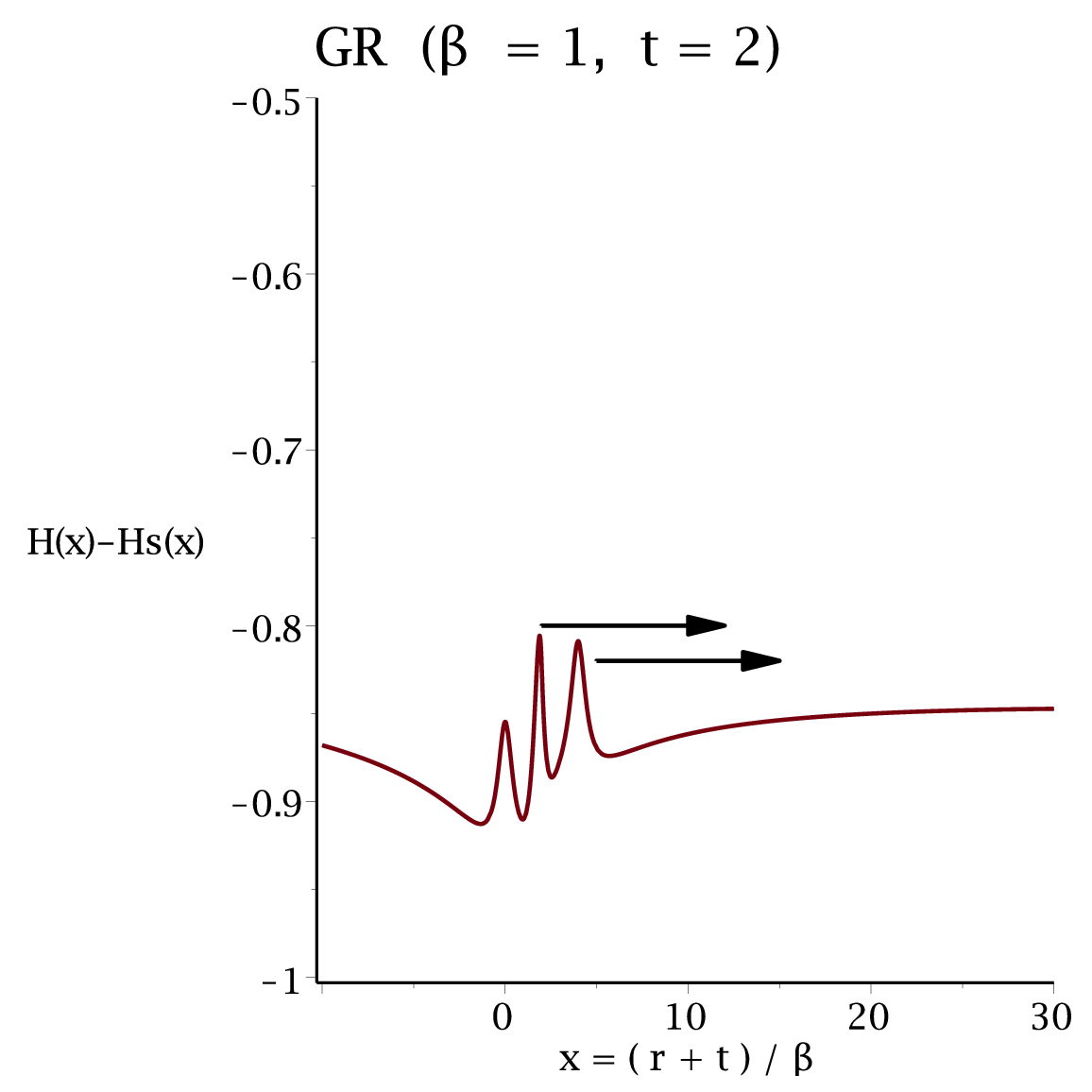}
	\includegraphics[width=7cm]{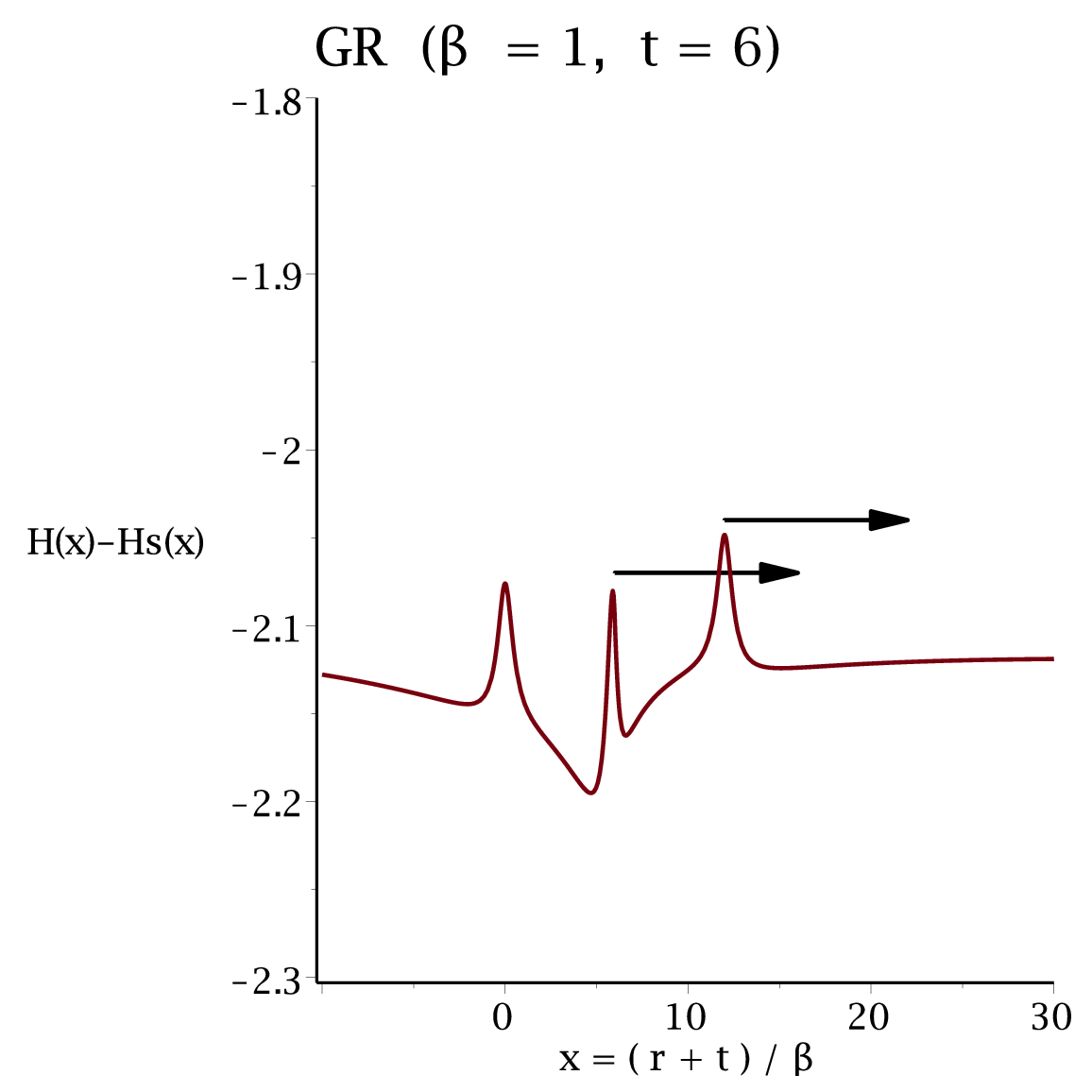}
	\includegraphics[width=7cm]{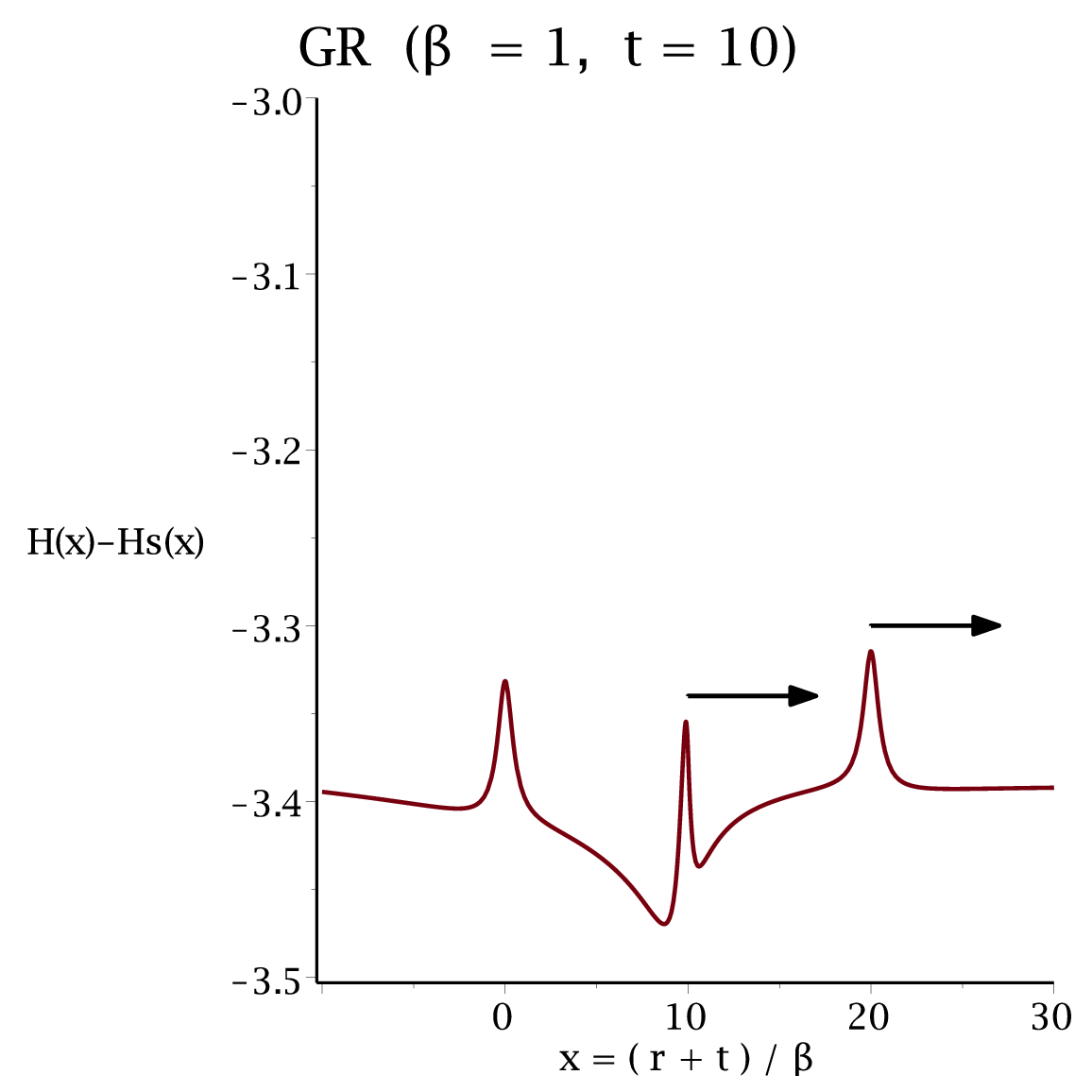}
	\caption{Wave component pulse $H$ of width $\beta$, using equation (\ref{Hrt}) and $\alpha=t$, $\beta=1$, $\gamma=r$ in equation (\ref{intpulse}), for four values of $t=0$, $t=2$, $t=6$ and $t=10$.} 
	\label{fig10}
\end{figure}

Substituting $\cos \left( {\mathrm w}t \right) ^{2}=[1+\cos \left( 2{\mathrm w}t \right)]/2$ and
$\cos \left( 2{\mathrm w}t \right)=[\exp\left( 2i{\mathrm w}t \right)+\exp\left( -2i{\mathrm w}t \right)]/2$
into equation (\ref{Hrtb}), we can integrate it and obtain it in terms of
the Meijer G function \cite{Prudnikov1990}, thus
\bqn
H(r,t)&=&-\frac{1}{2\beta \pi }\,
G^{2, 1}_{2, 2}\left( {\frac {{\beta}^{2}}{4{r}^{2}}}\, \Big\vert\,^{0, 1}_{\frac{1}{2}, \frac{1}{2}}\right)-\nb\\
&&\frac{1}{4\pi\left( -2\,it+\beta \right) }\,
G^{2, 1}_{2, 2}\left( {\frac { \left( -2\,it+\beta \right) ^{2}}{4{r}^{2}}}\, \Big\vert\,^{0, 1}_{\frac{1}{2}, \frac{1}{2}}\right)-\nb\\
&&\frac{1}{4 \pi \left( \beta+2\,it \right) }\,
G^{2, 1}_{2, 2}\left( {\frac { \left( \beta+2\,it \right) ^{2}}{4{r}^{2}}}\, \Big\vert\,^{0, 1}_{\frac{1}{2}, \frac{1}{2}}\right)-\nb\\
&&\frac{r}{{\beta}^2\pi} \left[ 
G^{2, 2}_{3, 3}\left( {\frac {{\beta}^{2}}{4{r}^{2}}}\, \Big\vert\,^{\frac{1}{2}, \frac{1}{2}, \frac{1}{2}}_{1, 0, \frac{3}{2}}\right)-
G^{2, 1}_{2, 2}\left( {\frac {{\beta}^{2}}{4{r}^{2}}}\, \Big\vert\,^{-\frac{1}{2}, \frac{1}{2}}_{1, 0}\right)
\right] 
\eqn
See Figures \ref{fig9} and \ref{fig10} for the pulse in GR. 

\section{Energy and Momentum of Cylindrical Gravitational Waves}

The gravitational energy-momentum pseudotensor, which Einstein initially formulated in 1916 following his discovery of the general relativity field equations, closely connects with the conventional canonical stress tensor applicable to matter fields in a flat spacetime. Many years ago, Scheidegger \cite{Scheidegger} questioned the well-defined existence of gravitational radiation. In response, Rosen \cite{Rosen:1956} explored the presence of energy and momentum in cylindrical gravitational waves. Employing the energy-momentum pseudotensors of Einstein and those of Landau and Lifshitz, Rosen calculated using cylindrical polar coordinates and determined that energy and momentum density components were null. These findings aligned with Scheidegger's suggestion that a physical system could not emit gravitational energy. However, Rosen \cite{Rosen:1958} later acknowledged an error and re-evaluated the calculations in cartesian coordinates, revealing nonvanishing and reasonable energy and momentum density. Despite Rosen's initial promise to publish the details elsewhere, this was not accomplished until later when he and another author \cite{Rosen:1993ea} re-examined the energy-momentum density components in cartesian coordinates following Einstein's prescription. Their subsequent findings were finite and reasonable \cite{Virbhadra:1995vu}. Here, we compute these quantities in EA theory using the equivalent expressions for the EA theory derived in reference \cite{Eling2006}.

The Einstein pseudotensor and the von Freud superpotential \cite{Julia1998} obey the following relations
\bqn
\partial_\mu t_\nu{}^\mu &=& 0 \label{Epseudo}.
\eqn
\bqn
t_\nu{}^\mu &=& -\partial_\gamma U_\nu{}^{\gamma\mu}\label{Esuper}.
\eqn
We present the pure Einstein part (GR) and pure aether contributions separately for easy comparison. 
\begin{eqnarray} ^{\rm Einstein}t^{\ \m}_\r &=& \frac{\sqrt{-g}}{16\pi G} \left(\d^{\
		\m}_\r (\G^\a_{\b\g}\G^\b_{\a\d}-\G^\a_{\a\b}\G^\b_{\g\d})g^{\g\d}
	\right. \nonumber\\&&
	\left.+\G^\b_{\r\a}\G^\g_{\g\b}g^{\m\a}-\G^\b_{\r\b}\G^\g_{\g\a}g^{\m\a}
	+\G^\a_{\r\a}\G^\m_{\b\g}g^{\b\g}\right. \nonumber\\&&
	\left.+\G^\m_{\r\a}\G^\b_{\b\g}g^{\a\g} -2
	\G^\m_{\a\b}\G^\a_{\r\g}g^{\b\g}\right), \label{Einsteinpsu}
\end{eqnarray}

\bea ^{\rm Aether}t_{\nu}{}^{\lambda} &=& \frac{1}{16\pi G}\left(2
{\sqrt{-g}} J^{\lambda}{}_{\rho}\nabla_{\nu} u^{\rho}
-\sqrt{-g}\{(J^{\lambda}{}_{\beta}+J_{\beta}{}^{\lambda})u^{\alpha}
\right. \nonumber\\
&&
\left.-(J^{\alpha}{}_{\beta}+J_{\beta}{}^{\alpha})u^{\lambda}+(J^{\lambda\alpha}
-J^{\alpha\lambda})u_{\beta}\}
\Gamma^{\beta}{}_{\alpha\nu}\right. \nonumber\\
&& \left.+ \delta_{\nu}^{\lambda} {\sqrt{-g}} L_{u}\right),
\label{aepsu} \eea

\beq ^{\rm Einstein}U_{\beta}{}^{\lambda\alpha} = \frac{1}{16\pi G}
\frac{1}{\sqrt{-g}}
g_{\beta\tau}\partial_\gamma\{(-g)(g^{\lambda\tau}g^{\a\gamma}-
g^{\alpha\tau}g^{\lambda\gamma})\}, \label{Einsteinsup}\eeq

and 

\bea ^{\rm Aether}U_\beta{}^{\lambda\alpha} &=& \frac{1}{16 \pi
	G}\sqrt{-g}
\left((J^{\lambda}{}_{\beta}+J_{\beta}{}^{\lambda})u^{\alpha}
\right. \nonumber\\
&& \left.
-(J^{\alpha}{}_{\beta}+J_{\beta}{}^{\alpha})u^{\lambda}+(J^{\lambda\alpha}
-J^{\alpha\lambda})u_{\beta}\right), \label{aesup}\eea

where $J^a{}_\b$ is defined in (\ref{Jdef}). The above decompositions of $t_\nu{}^\mu$  and $U_\nu{}^{\gamma\mu}$  into pure Einstein and aether pieces do not satisfy (\ref{Epseudo}) and (\ref{Esuper}) independently. When evaluating these pseudotensorial expressions, the metric must have connection coefficients vanish in  $O(1/r)$ or faster in the asymptotic limit in that coordinate system. If the coordinates are not chosen properly, these expressions will yield incorrect
energies and momenta. Further details on this can be found in \cite{Eling2006}.

Let us now calculate the energy-momentum pseudotensors given
by the equations (\ref{Einsteinpsu}) and (\ref{aepsu}) in 
cartesian coordinates, i.e., 
\bqn
ds^2=&&
{\frac {{e}^{-2\,\psi  } \left( {e}^{2\,H  }{x}^{2}+{y}^{2} \right)\, dx^2}{{x}^{2}+{y}^{2}}}+2\,{\frac {{e}^{-2\,\psi } \left( {e}^{2\,H  }-1 \right) xy\, dx\, dy}{{x}^{2}+{y}^{2}}}+\nb\\
&&{\frac {{e}^{-2\,\psi } \left( {e}^{2\,H  }{y}^{2}+{x}^{2}
		\right) \, dy^2}{{x}^{2}+{y}^{2}
}}+{e}^{2\,\psi }\,dz^2-{e}^{2\,H  -2\,\psi }\, d{t}^{2}.
\lb{ds2xy}
\eqn

In order to make the coordinate transformation from
the cylindrical to the cartesian coordinates, i.e.,
$\bar \psi(r, \theta, t) \rightarrow \psi \left( x,y,t \right)$
and
$\bar H(r,\theta,t) \rightarrow H \left( x,y,t \right)$,
let us use the usual polar coordinate transformation that says
$x(r,\theta)=r \cos(\theta)$
and
$y(r,\theta)=r \sin(\theta)$.
\footnote{
	Let us define now an arbitrary function $F$ in polar coordinates given by
	$\bar F\left( r,\theta \right) = F \left( x(r,\theta),\, y(r,\theta) \right)$.
	The chain rule says that $\bar F_r = F_x \, x_r + F_y \, y_r$ and $\bar F_\theta = F_x  \, x_\theta + F_y \, y_\theta$.
	In our case, we must impose that $\bar F_\theta=0$. Thus, we will be able to find the correct coordinate transformations, i.e.,
	$F_x = \bar F_r \cos(\theta)$; $F_y = \bar F_r \sin(\theta)$;
	$F_{xx} = \bar F_{rr} \cos(\theta)^2$; $F_{yy} = \bar F_{rr} \sin(\theta)^2$;
	$F_{xy} = \bar F_{rr} \cos(\theta) \sin(\theta)$; $F_{tx} = \bar F_{tr} \cos(\theta)$ and $F_{ty} = \bar F_{tr} \sin(\theta)$.}

The components of the energy-momentum pseudotensor are given
by the equation (\ref{Einsteinpsu}) are

\bqn
^{Einstein}t_x{}^x&=&
\frac{1}{8\pi G}\,\left[ {\frac { \left( x-y \right)  \left( y+x \right) {\psi_{{r}}}^{2}}{{r}^{2}}}+{{{\psi_{{t}}}^{2}}} \right],
\lb{texx}
\eqn

\bqn
^{Einstein}t_y{}^x&=&
{\frac {1}{4\pi G} \,\frac{{\psi_{{r}}}^{2}xy}{{r}^{2}}},
\eqn

\bqn
^{Einstein}t_x{}^y=
{\frac {1}{4\pi G} \,\frac{{\psi_{{r}}}^{2}xy}{{r}^{2}}},
\eqn

\bqn
^{Einstein}t_y{}^y&=&
-\frac{1}{8\pi G}\,\left[ {\frac { \left( x-y \right)  \left( y+x \right) {\psi_{{r}}}^{2}}
	{{r}^{2}}-{{\psi_{{t}}}^{2}}} \right],
\eqn

\bqn
^{Einstein}t_z{}^z&=&
-\frac{1}{8\pi G}\,\left[ {{e}^{2\,H}{\psi_{{r}}}^{2}+{e}^{2\,H}{\psi_{{t}}}^{2}-2\,
	{\psi_{{t}}}^{2}} \right],
\eqn

\bqn
^{Einstein}t_t{}^t&=&
-\frac{1}{8{\pi G}}\,{ \left( {\psi_{{r}}}^{2}+{\psi_{{t}}}^{2} \right) {e}^{2
		\,H}},
\lb{tett}
\eqn

\bqn
^{Einstein}t_x{}^t=
-\frac{1}{4\pi G}\,{\frac {x\psi_{{r}}\psi_{{t}}}{r}},
\eqn

\bqn
^{Einstein}t_t{}^x&=&
\frac{1}{4\pi G}\,{\frac {x\psi_{{r}}\psi_{{t}}{e}^{2\,H}}{r}},
\eqn

\bqn
^{Einstein}t_y{}^t&=&
-\frac{1}{4\pi G}\,{\frac {y\psi_{{r}}\psi_{{t}}}{r}},
\eqn

\bqn
^{Einstein}t_t{}^y&=&
\frac{1}{4\pi G}\,{\frac {y\psi_{{r}}\psi_{{t}}{e}^{2\,H}}{r}},
\eqn

\bqn
^{Einstein}t_z{}^t&=& ^{Einstein}t_t{}^z=0,
\lb{tetz}
\eqn
where the subscript notation means $F_v= \frac{\partial F}{\partial v}$
($F$ is an arbitrary function $F=F(r,t)$).

We can compare the components (\ref{tett})-(\ref{tetz}) with the papers 
\cite{Rosen:1993ea} and \cite{Virbhadra:1995vu}. We can notice that they
are the same unless the minus signal in all our components due to the different signature of the metric assumed in this work.

The components of the energy-momentum pseudotensor are given
by the equation (\ref{aepsu}) are

\bqn
^{Aether}t_x{}^x=
-\frac{1}{16\pi \,G}&&\left[ 
{\frac { \left( {r}^{2}{c_1}+{r}^{2}{c_4}-2\,{c_3}\,{x}^{2} \right) {H_{{r}}}^{2}}{{r}^{2}}}-\right.\nb\\
&&\left. {\frac { 2\left( {r}^{2}{c_1}+{r}^{2}{c_4}-2\,{c_3}\,{x}^{2} \right) \psi_{{r}}H_{{r}}}{{r}^{2}}}- \right. \nb\\
&&\left. {\left( {c_1}+{c_2}+{c_3} \right) {H_{{t}}}^{2}}+
{{2 \left( {c_1}+{c_2}+{c_3} \right) \psi_{{t}}H_{{t}}}}+\right. \nb\\
&&\left. {\frac { \left( {r}^{2}{c_1}+{r}^{2}{c_4}-2\,{c_3}\,{x}^{2} \right) {
			\psi_{{r}}}^{2}}{{r}^{2}}}-
{ \left( 3\,{c_1}+{c_2}+3\,{c_3} \right) {\psi_{{t}}}^{2}} \right],
\eqn

\bqn
^{Aether}t_y{}^x&=&
\frac{1}{8\pi G}\,{\frac {y{c_3}\,x \left( H_{{r}}-\psi_{{r}} \right) ^{2}}{
		{r}^{2}}},
\eqn

\bqn
^{Aether}t_x{}^y=
\frac{1}{8\pi G}\,{\frac {y{c_3}\,x \left( H_{{r}}-\psi_{{r}} \right) ^{2}}{
		{r}^{2}}},
\eqn

\bqn
^{Aether}t_y{}^y=
-\frac{1}{16{\pi G}}&& \left[ {\frac { \left( {r}^{2}{c_1}+{r}^{2}{c_4}-2\,{c_3}\,{y}
		^{2} \right) {H_{{r}}}^{2}}{{r}^{2}}}-\right.\nb\\
&&\left. {\frac {2 \left( {r}^{2}{c_1}+{r}^{2}{c_4}-
		2\,{c_3}\,{y}^{2} \right)\psi_{{r}}H_{{r}}}{{r}^{2}}}-\right.\nb\\
&&\left. { \left( {c_1}+{c_2}+{c_3} \right) {H_{{t}}}^{2}}+
{2 \left( {c_1}+{c_2}+{c_3} \right) \psi_{{t}}H_{{t}}}+ \right.\nb\\
&&\left. {\frac { \left( {r}^{2}{c_1}+{r}^{2}{c_4}-2\,{c_3}\,{y}^{2} \right) {\psi_{{r}}}^{2}}{{r}^{2}}}-{ \left( 3\,{c_1}+{c_2}+3\,{c_3} \right) {\psi_{{t}}}^{2}} \right],
\eqn

\bqn
^{Aether}t_z{}^z=
-\frac{1}{16\pi G}&&\left[ 
{ \left( {c_1}+{c_4} \right) {H_{{r}}}^{2}}-{ 2\left( {c_1}+{c_4} \right) \psi_{{r}}H_{{r}}}-{ \left( {c_1}+{c_2}+{c_3} \right) {H_
		{{t}}}^{2}}+ \right.\nb\\
&&\left. {2\left( {c_1}+{c_2}+{c_3}
	\right) \psi_{{t}}H_{{t}}}+{ \left( {c_1}+{
		c_4} \right) {\psi_{{r}}}^{2}}-{ \left( 3\,{
		c_1}+{c_2}+3\,{c_3} \right) {\psi_{{t}}}^{2}} \right],\nb\\
\eqn

\bqn
^{Aether}t_t{}^t=
-\frac{1}{16\pi G}&&\left[ 
{ \left( 3\,{H_{{r}}}^{2}-6\,\psi_{{r}}H_{{r}}+3\,{\psi_{
			{r}}}^{2}+3\,{\psi_{{t}}}^{2}-2\,H_{{t}}\psi_{{t}}+{H_{{t}}}^{2}
	\right) {c_1}}+\right.\nb\\
&&\left. { \left( H_{{t}}-\psi_{{t}} \right) ^{2}{c_2}}-\right.\nb\\
&&\left. { \left( 2\,{H_{{r}}}^{2}-4\,\psi_{{r}}H_{{r}}+2\,{\psi_{{r}}}^{2}-3\,{\psi_{{t}}}^{2}+2\,H_{{t}}\psi_{{t}}-{H_{{t}}}^{2} \right) {c_3}}+\right.\nb\\
&&\left. {3 \left( H_{{r}}-\psi_{{r}} \right) ^{2}{c_4}} \right],
\eqn

\bqn
^{Aether}t_x{}^t=
-\frac{1}{8\pi G}\,&&\left[ 
{\frac {x\psi_{{t}} \left( 2\,r\psi_{{r}}+{e}^{2\,H}-1 \right) {
			c_1}}{{r}^{2}}}-\right. \nb\\
&&\left. \frac {x \left( H_{{t}}-\psi_{{t}}
	\right)  \left( -r\psi_{{r}}+H_{{r}}r+{e}^{2\,H}-1 \right) {c_2}}{{r}^{2}}+\right.\nb\\
&&\left. {\frac {x\psi_{{t}} \left( 2\,r\psi_{{r}}+{e}^{2
			\,H}-1 \right) {c_3}}{{r}^{2}}}+\right.\nb\\
&&\left. {\frac {x \left( H_{{t}
		}-\psi_{{t}} \right)  \left( H_{{r}}-\psi_{{r}} \right) {c_4}}{r
}} \right],
\eqn

\bqn
^{Aether}t_t{}^x=
-\frac{1}{8\pi G}&&\left[ {\frac {x \left( H_{{t}}-\psi_{{t}} \right)  \left( H_{{r}}-\psi_{{r}} \right)  \left( 2\,{c_1}-2\,{c_2}-{c_3} \right) }{r}}
\right],
\eqn

\bqn
^{Aether}t_y{}^t=\frac{1}{8\pi G}&&\left[ 
-{\frac {y\psi_{{t}} \left( 2\,r\psi_{{r}}+{e}^{2\,H}-1 \right) {c_1}}{{r}^{2}}}+\right.\nb\\
&&\left. {\frac {y \left( H_{{t}}-\psi_{{t}}
		\right)  \left( -r\psi_{{r}}+H_{{r}}r+{e}^{2\,H}-1 \right) {c_2}}{
		{r}^{2}}}-\right.\nb\\
&&\left. \frac {y\psi_{{t}} \left( 2\,r\psi_{{r}}+{e}^{2
		\,H}-1 \right) {c_3}}{{r}^{2}}-\right.\nb\\
&&\left. {\frac {y \left( H_{{t}}-\psi_{{t}} \right)  \left( H_{{r}}-\psi_{{r}} \right) {c_4}}{r
}}\right],
\eqn

\bqn
^{Aether}t_t{}^y&=&
\frac{1}{8\pi G}\,{\frac {y \left( 2 {c_1} + 2 {c_2} + {c_3}  \right) 
		\left( H_{{t}}H_{{r}}-\psi_{{t}}H_{{r}}-H_{{t}}\psi_{{r}}+\psi_{{t}}
		\psi_{{r}} \right) }{r}},
\eqn

\bqn
^{Aether}t_z{}^t&=& ^{Aether}t_t{}^z=0,
\lb{taezz}
\eqn

As said before, when
evaluating the energy-momentum pseudotensor, the
metric must be written in a coordinate system where the
connection coefficients vanish like $\mathcal{O}(1/r)$ or faster in the
asymptotic limit. Thus, 
we can see from equations (\ref{texx})-(\ref{taezz}) that the energy and momentum 
densities are finite and reasonable at infinity. From equations (\ref{Psirt}) and (\ref{Hrt}), we can notice that the quantities $H_t$ and $\psi_t$ are
oscillatory functions of the time. Since the quantities $H_r$ and $\psi_r$ are also oscillatory in the radial coordinate and decreasing at infinity. Thus, we have finite values of the energy-momentum pseudotensors
at infinity.

We can notice that EA energy-momentum pseudotensors are completely different from the GR ones by construction, i.e., if we
	assume $c_1=c_2=c_3=c_4=$ all the EA energy-momentum pseudotensor components
	are zero.

\section{Conclusions}

We have obtained the gravitational wave equation for cylindrical symmetry in the EA theory. This is the generalization of the Einstein-Rosen wave equation in GR. We have shown that the gravitational wave in the EA is periodic in time, both for $\Psi(r,t)$ and $H(r,t)$ metric functions. However, in GR, $\Psi(r,t)$ is periodic in time, but $H(r,t)$ is semi-periodic in time, having a secular drifting in the wave frequency. Due to this frequency drifting, the evolution of wave pulses of a given width emitted is entirely different in both theories in the $H(r,t)$ metric 
function. Another difference between the two theories is the gravitational wave velocity. While in GR, the waves propagate with the speed of light, 
in EA, there is not an upper limit in the wave velocity, reaching infinity value if $c_{13} \rightarrow 1$ and zero if $c_{13} \rightarrow -\infty$. Also, aether has contributions to energy-momentum pseudotensor and superpotential. In principle, all these characteristics could be a way to differentiate  GR and EA theories observationally.

\section {Acknowledgments}

The author (RC) acknowledges the financial support from FAPERJ (E-26/171.754/2000, E-26/171.533/2002 and E-26/170.951/2006). MFAdaS  acknowledges the financial support from CNPq-Brazil, FINEP-Brazil (Ref. 2399/03), FAPERJ/UERJ (307935/2018-3), { FAPERJ (E-26/211.906/2021)} and from CAPES (CAPES-PRINT 41/2017).

\section{References}

\end{document}